\documentclass[lettersize,journal]{IEEEtran}

 \expandafter\def\expandafter\normalsize\expandafter{%
     \normalsize%
     \setlength\abovedisplayskip{4pt}%
     \setlength\belowdisplayskip{4pt}%
     \setlength\abovedisplayshortskip{2pt}%
     \setlength\belowdisplayshortskip{2pt}%
 }

\usepackage{cite}
\usepackage{amsmath,amssymb,amsfonts,empheq}
\usepackage{graphicx}
\usepackage{textcomp}
\usepackage{enumerate}
\usepackage{stfloats}
\usepackage{multirow}
\usepackage{booktabs}
\usepackage{color}
\usepackage{array}
\usepackage{makecell}
\usepackage{url}
\usepackage{booktabs}
\usepackage{etoolbox}
\usepackage{epstopdf}
\usepackage[section]{placeins}
\usepackage{float}
\usepackage{colortbl}
\usepackage{listings}
\usepackage{algorithm}
\usepackage{algpseudocode}
\floatname{algorithm}{Algorithm}
\usepackage{epstopdf}
\usepackage{setspace}
\usepackage{bm}
\usepackage{amsmath}

\usepackage{subcaption}

\allowdisplaybreaks[4]

\makeatletter
\renewcommand{\maketag@@@}[1]{\hbox{\m@th\normalsize\normalfont#1}}%
\makeatother

\makeatletter
\patchcmd{\@makecaption}
  {\scshape}
  {}
  {}
  {}
\makeatother

\begin{document}

\title{\textcolor{black}{Wideband Near-Field Sensing in ISAC: Unified Algorithm Design and Decoupled Effect Analysis}}

\author{Ruiyun Zhang, Zhaolin Wang, Zhiqing Wei, Yuanwei Liu, Zehui Xiong, Zhiyong Feng %

\thanks{Ruiyun Zhang, Zhiqing Wei, and Zhiyong Feng are with the Key Laboratory of Universal Wireless
Communications, Ministry of Education, School of Information and Communication Engineering, Beijing University of Posts and Telecommunications, Beijing 100876, China. (e-mail: zhangruiyun@bupt.edu.cn; weizhiqing@bupt.edu.cn; fengzy@bupt.edu.cn).}%
\thanks{Zhaolin Wang and Yuanwei Liu are with the Department of  Electrical and Computer Engineering, The University of Hong Kong,  Hong Kong.
 (e-mail: zhaolin.wang@hku.hk; yuanwei@hku.hk).}%
\thanks{Zehui Xiong is the School of Electronics, Electrical Engineering and
Computer Science, Queen’s University Belfast,  United Kingdom. (e-mail: z.xiong@qub.ac.uk).}%
\thanks{Zhiqing Wei is the corresponding author of this article.}
}



\IEEEaftertitletext{\vspace{-2.5\baselineskip}} 
\maketitle

\begin{abstract}
To advance integrated sensing and communications (ISAC) in sixth-generation (6G) extremely large-scale multiple-input multiple-output (XL-MIMO) networks, a low-complexity compressed sensing (CS)-based dictionary design is proposed for wideband near-field (WB-NF) target localization. Currently, the massive signal dimensions in the WB-NF regime impose severe computational burdens and high spatial-frequency coherence on conventional grid-based algorithms. Furthermore, a unified framework exploiting both wideband (WB) and near-field (NF) effects is lacking, and the analytical conditions for simplifying this model into decoupled approximations remain uncharacterized. To address these challenges, the proposed algorithm mathematically decouples the mutual coherence function and introduces a novel angle-distance sampling grid with customized distance adjustments, drastically reducing dictionary dimensions while ensuring low coherence. To isolate the individual WB and NF impacts, two coherence-based metrics are formulated to establish the effective boundaries of the narrowband near-field (NB-NF) and wideband far-field (WB-FF) regions, where respective multiple signal classification (MUSIC) algorithms are utilized. Simulations demonstrate that the CS-based method achieves robust performance across the entire regime, and the established boundaries provide crucial theoretical guidelines for WB and NF effect decoupling.

\end{abstract}


\IEEEpeerreviewmaketitle
\vspace{-10pt}
\section{Introduction}
\IEEEPARstart{I}{ntegrated} sensing and communications (ISAC) is envisioned as a defining technology for sixth-generation (6G) wireless networks~\cite{liu2022integrated, zhang20196g, wei2024deep}. Driven by the requirements for high-resolution sensing and massive data transmission, there is a paradigm shift from fifth-generation (5G) massive multiple-input multiple-output (MIMO)~\cite{bjornson2023twenty, bjornson2014massive,10255745} to extremely large-scale MIMO (XL-MIMO)~\cite{10379539, wang2025analytical} for 6G. Coupled with the use of high-frequency bands such as millimeter-wave (mmWave) and terahertz (THz), the drastic expansion of the large aperture fundamentally extends the Rayleigh distance, rendering the conventional far-field (FF) planar wavefront model inapplicable and necessitating the rigorous treatment of non-negligible near-field (NF) effects. Simultaneously, ultra-wideband (WB) transmissions become indispensable to provide abundant frequency-domain resources for delay-based ranging accuracy. Consequently, sensing in 6G will inevitably operate in a wideband near-field (WB-NF) regime\cite{10663786}.

In the WB-NF regime, the target distance information is embedded in both the spatial and frequency domains through distinct physical mechanisms. Historically, target sensing in ISAC systems was extensively investigated within the wideband far-field (WB-FF) framework~\cite{10736660, 10851319, xu2025does, 10634583, 10285442}. These approaches effectively decoupled spatial and temporal processing, resolving angles via the array response, while deriving distances from the WB frequencies. For large-aperture arrays envisioned in 6G systems, this planar wavefront assumption leads to model mismatch and spatial defocusing in the extended NF region. Furthermore, the spherical wavefront introduces a non-linear quadratic phase variation, which enables distance estimation via the spatial array response, alleviating the necessity for extensive system bandwidth~\cite{10663786}. Driven by these two factors, most recent research efforts initially focused on narrowband near-field (NB-NF) sensing~\cite{10379539, 10663521, friedlander2019localization, 11030222, 10388218, wang2023near, zuo2018subspace, 11078780}. For instance, pioneering tutorials and overview works~\cite{10663521, 10379539, friedlander2019localization} systematically articulated the paradigm shift, fundamental principles, and inherent opportunities of NF ISAC introduced by XL-MIMO architectures. From a theoretical performance perspective, fundamental analytical bounds, such as the Cramér-Rao bound (CRB), were rigorously derived to quantify and analyze the performance enabled by the spherical wavefront curvature~\cite{11030222, 10388218, wang2023near_a}. Building upon these theoretical foundations, a variety of parameter estimation algorithms were developed. Notably, subspace fitting-based~\cite{zuo2018subspace} and modified multiple signal classification (MUSIC) approaches~\cite{11078780, friedlander2019localization} were proposed to realize high-resolution target localization by exploiting the distance-dependent non-linear phase structures. While these NB-NF approaches achieve distance estimation via wavefront curvature, this focusing gain diminishes rapidly with distance, necessitating WB signals for robust delay-based sensing. However, directly applying the NB-NF assumption to WB-NF systems inherently ignores the substantial beam squint effect, which causes severe energy defocusing, ultimately nullifying the NF curvature advantages.


Therefore, a unified WB-NF algorithm is necessary to seamlessly bridge these regimes. Recognizing this necessity, emerging studies have recently ventured into WB-NF sensing, which can be broadly categorized into two streams. The first stream focuses on fundamental formulation and theoretical limits, establishing analytical channel models and performance bounds~\cite{10663786, 10663521, zhao2024modeling, wang2025performance, 10934783, 10521567}. The second stream centers on target localization and channel estimation\cite{cui2022channel, 10271123}. For instance, the authors in \cite{cui2022channel} pioneered compressed sensing (CS)-based NF channel estimation by exploiting the polar-domain sparsity. Furthermore, the work in \cite{10271123} introduced true-time-delay (TTD) to control the beam squint trajectory, inversely exploiting beam squint for efficient target localization.

Despite these pioneering efforts, a unified estimation algorithm capable of jointly exploiting the NF spherical curvature effect and the frequency-dependent WB effect without relying on customized hardware modifications (e.g., TTDs) is still lacking. Specifically, WB-NF sensing introduces a massive volume of signal data across extensive subcarriers and large-scale antennas, which poses a formidable computational burden. Fortunately, since the number of targets is typically very small relative to the spatial-frequency dimensions, the target parameters exhibit inherent sparsity. This physical characteristic makes CS an ideal foundational framework to achieve high-resolution estimation with drastically reduced computational overhead. However, directly applying CS to the unified WB-NF regime is non-trivial, since conventional grid-based algorithms suffer from severe spatial-frequency coherence among dictionary atoms, which degrades the sparse recovery performance. More fundamentally, the physical mechanisms dictating when the unified WB-NF model degrades into decoupled approximations (i.e., the NB-NF and WB-FF) remain uncharacterized. Driven by the above considerations and the inevitable trend towards the ISAC paradigm, this paper investigates the performance of target sensing within a WB-NF communication system, focusing on the unified algorithm design and decoupled NF and WB effect analysis. The main contributions are summarized as follows:
\begin{itemize}
    \item A low-complexity CS-based dictionary design algorithm is proposed for target localization in WB-NF systems. By decoupling the mutual coherence function, the proposed algorithm utilizes a novel angle-distance sampling grid with a customized distance adjustment strategy, which significantly reduces dictionary dimensions while rigorously guaranteeing low mutual coherence.

    \item To isolate the individual impacts of the WB and NF effects, two coherence functions are formulated to establish the boundaries of the NB-NF and WB-FF regions. Within these regions, decoupled MUSIC algorithms are developed to achieve high-resolution target localization.

    \item Extensive simulations are conducted to validate the robust performance of the proposed CS-based WB-NF localization method across the entire sensing region. Furthermore, the results demonstrate that the established NB-NF and WB-FF boundaries provide crucial guidelines for decoupling the WB and NF effects in ISAC systems.
\end{itemize}

\begin{figure}[t!]
\centering
\includegraphics[width=0.70\linewidth]{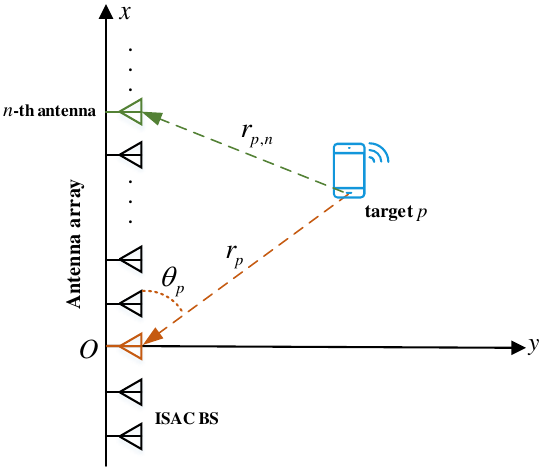}
\caption{Illustration of the considered NF system model.}
\label{fig:fig-1}
\end{figure}

The remainder of this paper is organized as follows. Section II introduces the system and signal model for WB-NF sensing. Section III presents the sparse matrix formulation and the proposed low-coherence dictionary design. Section IV investigates the analytical characterization of the WB and NF effects and proposes two MUSIC benchmarks. Simulation results are provided in Section V. Finally, Section VI concludes the paper.

\emph{Notations:} Scalars, vectors, and matrices are denoted by non-boldface, boldface lowercase, and boldface uppercase letters, respectively. The space of $N_c \times N_s$ complex matrices is denoted by $\mathbb{C}^{N_c \times N_s}$. The superscripts $(\cdot)^T$ and $(\cdot)^H$ denote the transpose and conjugate transpose operations, respectively. The operator $\mathrm{diag}(\mathbf{x})$ represents a diagonal matrix with the elements of vector $\mathbf{x}$ on its main diagonal. The operators $\lfloor \cdot \rfloor$ and $\lceil \cdot \rceil$ denote the floor and ceiling of their arguments, respectively. Calligraphic uppercase letters (e.g., $\mathcal{S}$) are used to denote sets, and $|\mathcal{S}|$ denotes the cardinality of the set $\mathcal{S}$. The hat operator $\hat{(\cdot)}$ denotes the estimated value of a variable. Finally, $\mathcal{CN}(\boldsymbol{\mu}, \mathbf{\Sigma})$ denotes the circularly symmetric complex Gaussian distribution with mean vector $\boldsymbol{\mu}$ and covariance matrix $\mathbf{\Sigma}$, and $\mathbf{I}_N$ is the $N \times N$ identity matrix.

\vspace{-5pt}
\section{System Model}
\label{sec:sig}
We consider an orthogonal frequency-division multiplexing (OFDM) ISAC system in the NF region. As illustrated in Fig.~\ref{fig:fig-1}, the base station (BS) is equipped with an $N$-element uniform linear array (ULA) deployed along the $x$-axis, and $P$ single-antenna active targets located in the NF region. The center of the ULA is aligned with the origin of the Cartesian coordinate system, and the antenna elements are spaced by $d = \lambda_c / 2$, where $\lambda_c$ represents the central carrier wavelength. Specifically, the Cartesian coordinates of the $n$-th antenna element are expressed as $(\delta_n d, 0)$, where $\delta_n = n - (N - 1)/2$ represents the symmetric element index for $n = 0, \dots, N - 1$.

The targets are located within the Fresnel region, strictly upper-bounded by the Rayleigh distance $R_{r} = 2D^2/\lambda_c$, where $D = (N - 1)d$ denotes the array aperture. We assume a quasi-static scenario where the position of the $p$-th target, denoted by the coordinate $(r_p, \theta_p)$ in polar form, remains constant within a coherent processing interval (CPI). Here, $r_p$ represents the distance from the origin, and $\theta_p \in [0, \pi]$ denotes the angle relative to the $x$-axis. Consequently, the distance between the $p$-th target and the $n$-th antenna element is derived as
\begin{equation}
\label{eq:eq-1}
{r_{p,n}} = \sqrt {{r_p^2} + \delta _n^2{d^2} - 2r_p{\delta _n}d\cos \theta_p }.
\end{equation}
Focusing on the dominant Line-of-Sight (LoS) propagation path in the NF environment, the propagation delay is given by $\tau_{p,n} = r_{p,n} / c$, with $c$ being the speed of light.

\vspace{-5pt}
\subsection{Transmit Signal Model}
The uplink transmission utilizes OFDM frames consisting of $K$ symbols with $M$ subcarriers. The total bandwidth is $B = M \Delta f$, where $\Delta f$ is the subcarrier spacing. Correspondingly, the effective symbol duration is given by $T_d = 1/\Delta f$. The complex baseband signal transmitted by the $p$-th target is~\cite{10887024}
\begin{equation}
\label{eq:eq-2}
{\tilde{x}_p}(t) = \dfrac{1}{{\sqrt {{M}} }} \sum\limits_{k = 0}^{{K} - 1} \sum\limits_{m = 0}^{{M} - 1}  {{{s}}_{k,m,p}}{e^{\mathrm{j} 2\pi m\Delta f (t - k T_t) }}
{\mathrm{rect}} \left( \frac{{t - k{T_t}}}{{{T_t}}} \right),
\end{equation}
where $T_t = T_d + T_g$ is the total OFDM symbol duration including the effective duration $T_d$ and cyclic prefix (CP) $T_g$. The term $s_{k,m,p}$ denotes the pilot symbol transmitted on the $m$-th subcarrier of the $k$-th OFDM symbol by the $p$-th target, and ${\mathrm{rect}}( \cdot )$ is the rectangular function, defined as
\begin{equation}
\label{eq:eq-3}
\mathrm{rect} (u) = \begin{cases}
1, & 0 \le u < 1, \\
0, & \text{otherwise}.
\end{cases}\vspace{-10pt}
\end{equation}

\subsection{Received Signal Model}
The received signal at the $n$-th antenna is a superposition of signals from $P$ targets, distorted by delay and additive noise
\begin{equation}
\label{eq:eq-4}
y_{n}(t) = \sum\limits_{p = 1}^P \beta _{p,n} \, \tilde x_p(t - \tau_{p,n}) \, e^{ - \mathrm{j}2\pi f_c \tau_{p,n}} + w_{n}(t),
\end{equation}
where $w_n(t) \sim \mathcal{CN}(0, \sigma^2)$ represents the additive white Gaussian noise (AWGN). As mentioned above, the propagation delay between the $p$-th target and the $n$-th BS antenna is defined strictly by the geometric distance as $\tau_{p,n} = r_{p,n} / c$. The path loss coefficient is given by $\beta_{p,n} \approx \frac{c}{4\pi f_c r_{p,n}}$~\cite{10271123}.

Following CP removal, the continuous-time received signal $y_n(t)$ is sampled at discrete instants $t = k T_t + T_g + i T_s$, where $i = 0, \dots, M-1$ denotes the fast-time sample index. The sampling interval is set to $T_s = 1/B = 1/(M \Delta f)$, adhering to the Nyquist criterion $T_d = M T_s$. For canonical derivation purposes, we initially focus on the noiseless signal component. Assuming perfect time synchronization, the resulting discrete-time signal at the $n$-th antenna is derived as
\begin{equation}
\label{eq:eq-5}
\begin{aligned}
&{y_n}(k{T_t} + T_g + i{T_s}) \triangleq {y_{k, n}}\left[ i \right]\\
&\approx \frac{1}{{\sqrt {M} }} {\sum\limits_{p = 1}^P \sum\limits_{m = 0}^{M - 1} {{\beta _{p}}} {s_{k,m,p}}\,{e^{\mathrm{j} 2\pi \frac{{mi}}{M}}}{e^{ - \mathrm{j} 2\pi {f_m}{\tau_{p,n}}}}}.
\end{aligned}
\end{equation}
When expanding the received signal over subcarriers, we rigorously account for the path loss by initially defining the amplitude as $\beta_{p,n,m} \triangleq \frac{c}{4\pi f_m r_{p,n}}$, where $f_m = f_c + m\Delta f$. In the last step of Eq.~(\ref{eq:eq-5}), this varying amplitude is approximated as follows. First, given that the transmission distance exceeds the array aperture ($r_p > D$), the amplitude variation across array elements is negligible compared to the phase sensitivity, yielding $\beta_{p,n,m} \approx \beta_{p,m} = \frac{c}{4\pi f_m r_p}$~\cite{wang2023near}. Second, under the narrowband (NB) amplitude assumption ($f_c \gg B$), the slow $1/f_m$ decay permits the approximation $\beta_{p,m} \approx \beta_p = \frac{c}{4\pi f_c r_p}$~\cite{11005389, 11357477}. Note that while the amplitude is treated as constant, the exact phase progression $e^{-\mathrm{j}2\pi f_m \tau_{p,n}}$ is strictly retained to capture the WB effect and NF characteristics.

By stacking the received signals across the $N$ antenna elements, we define the spatial snapshot vector at time index $i$ as $\mathbf{y}_k[i] \triangleq [y_{k,0}[i], \dots, y_{k,N-1}[i]]^{\mathrm{T}} \in \mathbb{C}^{N \times 1}$. Substituting the approximation Eq.~\eqref{eq:eq-5}, the received signal vector is given by
\begin{equation}
\label{eq:eq-6}
\mathbf{y}_k[i] \approx \sqrt{\frac{1}{M}} \sum_{p=1}^{P} \beta_p \sum_{m=0}^{M-1} \mathbf{a}_m(r_p, \theta_p) s_{k,m,p} e^{\mathrm{j} 2\pi \frac{mi}{M}},
\end{equation}
where the term $\mathbf{a}_m(r_p, \theta_p) \in \mathbb{C}^{N \times 1}$ denotes the NF steering vector for the $m$-th subcarrier, defined as $\mathbf{a}_m(r_p, \theta_p) = \left[ e^{-\mathrm{j} 2\pi f_m \tau_{p,0}}, \dots, e^{-\mathrm{j}2\pi f_m \tau_{p,{N-1}}} \right]^{\mathrm{T}}$. To facilitate parameter estimation during the sensing interval, we utilize reference pilot symbols that are constant across subcarriers for the $k$-th OFDM symbol, i.e., $s_{k, 0, p} = \dots = s_{k, M-1, p} = s_{k,p}$. Applying the unitary Discrete Fourier Transform (DFT) to the time-domain snapshots $\mathbf{y}_k[i]$, and explicitly reintroducing the additive noise component to complete the statistical model, we obtain the frequency-domain signal at the $m$-th subcarrier
\begin{equation}
\label{eq:eq-7}
\begin{aligned}
\mathbf{y}_{k, m} \!\!&=\!\! \mathrm{DFT}\left\{ \mathbf{y}_k[i] \right\}_m \!\!=\!\! \sum_{p=1}^{P} \beta_p s_{k, p} \mathbf{a}_m(r_p, \theta_p) + \mathbf{w}_{k, m},
\end{aligned}
\end{equation}
where $\mathbf{w}_{k, m} \sim \mathcal{CN}(\mathbf{0}, \sigma^2 \mathbf{I}_N)$ represents the noise vector, preserving the white statistics under unitary transformation.

Following that, by stacking the observations from all $M$ subcarriers, we formulate the sensing model for the $k$-th symbol as $\mathbf{y}_k = [\mathbf{y}_{k, 0}^{\mathrm{T}}, \dots, \mathbf{y}_{k, M-1}^{\mathrm{T}}]^{\mathrm{T}}$, which is expressed as
\begin{equation}
\label{eq:eq-8}
\mathbf{y}_k = \sum_{p=1}^{P} \beta_p s_{k,p} \mathbf{a}(r_p, \theta_p) + \mathbf{w}_k,
\end{equation}
where $\mathbf{w}_k = [\mathbf{w}_{k, 0}^{\mathrm{T}}, \dots, \mathbf{w}_{k, M-1}^{\mathrm{T}}]^{\mathrm{T}}$ is the WB noise vector. The term $\mathbf{a}(r_p, \theta_p) \in \mathbb{C}^{NM \times 1}$ denotes the joint spatial-frequency steering vector, defined as $\mathbf{a}(r_p, \theta_p) = [\mathbf{a}_0^{\mathrm{T}}(r_p, \theta_p), \dots, \mathbf{a}_{M-1}^{\mathrm{T}}(r_p, \theta_p)]^{\mathrm{T}}$.

Subsequently, the received signal matrix over $K$ symbols, denoted as $\mathbf{Y} = [\mathbf{y}_0, \dots, \mathbf{y}_{K-1}] \in \mathbb{C}^{NM \times K}$, is formulated as
\begin{equation}
\label{eq:eq-9}
\mathbf{Y} = \sum_{p=1}^{P} \beta_p \mathbf{a}(r_p, \theta_p) \mathbf{s}_p^{\mathrm{T}} + \mathbf{W},
\end{equation}
where $\mathbf{s}_p = [s_{0,p}, \dots, s_{K-1,p}]^{\mathrm{T}} \in \mathbb{C}^{K \times 1}$ represents the pilot sequence vector for the $p$-th target, and $\mathbf{W} = [\mathbf{w}_{0}, \dots, \mathbf{w}_{K-1}] \in \mathbb{C}^{NM \times K}$ is the aggregated noise matrix.

\subsection{Problem Statement}
The objective of this work is to accurately estimate the distance and angle parameters $(r_p, \theta_p)$ for all $P$ sensing targets from the multi-carrier observation matrix $\mathbf{Y}$. Based on the unified model in \eqref{eq:eq-9}, the core of the sensing task lies in the characterization of the joint spatial-frequency steering vector $\mathbf{a}(r, \theta)$, which presents twofold technical challenges.

First, the target information is embedded in a dual-domain coupled phase structure. Specifically, the NF effect introduces a non-linear phase curvature across the large-scale array (the $n$-dimension), while the WB effect manifests as a frequency-dependent phase progression across subcarriers (the $m$-dimension). Unlike conventional models, the distance $r$ here serves as a common factor influencing both the spherical wavefront shape and the time-of-arrival delay, leading to a high-dimensional parameter coupling that invalidates simplified decoupled estimation. Second, extracting the distance and angle parameters from $\mathbf{a}(r, \theta)$ encounters a formidable computational bottleneck. The aggregated dimension $NM$ in modern XL-MIMO systems leads to a prohibitive cost for classical grid-based methods. Furthermore, when formulating this as a sparse recovery problem, the joint WB-NF manifold exhibits high mutual coherence between neighboring atoms in the distance-angle grid, which necessitates a sophisticated dictionary design method to ensure robust recovery performance with reduced complexity.

\section{Sparse Matrix Formulation and Localization Algorithm Design}
\label{sec:est}
In this section, we detail the proposed low-complexity WB-NF localization framework. We first formulate a sparse matrix representation and analytically decouple the mutual coherence function to facilitate efficient dictionary design. Subsequently, we develop a novel angle-distance sampling grid that incorporates Fresnel-based spatial analysis and a bandwidth-constrained alignment strategy, effectively reducing the dictionary dimensionality without compromising the mutual coherence properties. The section concludes with the presentation of a unified CS-based algorithm for robust target parameter estimation across the entire WB-NF regime.

\vspace{-4pt}
\subsection{Sparse Matrix Representation for WB-NF Sensing}
Based on the signal model derived in Eq.~\eqref{eq:eq-9}, the WB-NF channel vector $\mathbf{h} \in \mathbb{C}^{NM \times 1}$ is theoretically defined as the superposition of the steering vectors from all $P$ targets
\begin{equation}
\label{eq:eq-10}
\mathbf{h} = \sum_{p=1}^{P} \beta_p \mathbf{a}(r_p, \theta_p),
\end{equation}which characterizes the joint space-frequency response, incorporating both the path gains and the geometric information of the targets. Our objective is to estimate the target parameters embedded in $\mathbf{h}$ from the noisy observation $\mathbf{Y}$.

To facilitate employing CS techniques, we adopt a grid-based sparse representation. Extending the angle-distance framework~\cite{cui2022channel}, and assuming the targets fall on the sampling grid, the linear sparse model is formulated as
\begin{equation}
\label{eq:eq-11}
\mathbf{h} = \mathbf{B} \mathbf{h}_s,
\end{equation}
where the matrix ${{\mathbf{B}}} \in {{\mathbb C}^{NM \times Q_{\theta}Q_r}}$ is the overcomplete dictionary whose columns are sampled steering vectors $\mathbf{a}(r_{i},\theta_{j})$, and ${{\mathbf{h}}_{s}} \in \mathbb{C}^{Q_{\theta}Q_r \times 1}$ is a sparse coefficient vector  indicating the locations and strengths of the targets, with at most $P$ significant nonzero entries (typically $P \ll Q_{\theta}Q_r$) corresponding to the dominant LoS components. Here, $Q_{\theta}Q_r$ gives the total number of sampled NF steering vectors in the angle-distance domain, with $Q_r$ and $Q_{\theta}$ representing the numbers of distance and angular sampling points, respectively. Consequently, a key challenge is designing the angle-distance sampling strategy for the dictionary matrix ${\mathbf{B}}$ to strictly satisfy the restricted isometry property (RIP) or simplify the coherence structure~\cite{5967912}.

To enable precise target identification and parameter estimation within the CS framework~\cite{cui2022channel}, the angular and distance sampling must be meticulously designed to ensure that the dictionary matrix ${\mathbf{B}}$ achieves minimal mutual coherence among its columns. The mutual coherence, defined as
\begin{equation}
\label{eq:eq-12}
\begin{aligned}
g(r_{i}, \theta_{i}, r_{j}, \theta_{j}) &= \left| \mathbf{a}^H(r_{i}, \theta_{i}) \mathbf{a}(r_{j}, \theta_{j}) \right| \\
&= \frac{1}{NM} \left| \sum_{m=0}^{M-1} \sum_{n=0}^{N-1} e^{\mathrm{j} k_m (r_{i,n} - r_{j,n})} \right|,
\end{aligned}
\end{equation}
where the indices $i, j \in \{1, \dots, Q_{\theta}Q_r\}$ denote two arbitrary columns in the dictionary, associated with the spatial grid points $(r_{i}, \theta_{i})$ and $(r_{j}, \theta_{j})$, respectively. The wave number is defined as ${k_m} = \frac{{2\pi {f_m}}}{c} = \frac{{2\pi }}{{{\lambda _m}}}$, corresponding to the $m$-th carrier frequency. Consequently, the global mutual coherence of the dictionary is determined by $\mu({\mathbf B})=\max_{i \neq j} g(r_{i},\theta_{i},r_{j},\theta_{j})$. Since all dictionary columns are $\ell_2$-normalized to unit norm (implying each entry has a modulus of $1/\sqrt{NM}$), both the pairwise coherence $g(\cdot)$ and the global coherence $\mu(\mathbf B)$ are strictly bounded within the interval $[0,1]$.

Considering a ULA configuration with $N = 2 N' + 1$ antennas, the relative element index ${\delta _n}$, as defined in Section~\ref{sec:sig}, is given by ${\delta _n} = n - N'$. To simplify the distance model in Eq.~\eqref{eq:eq-1}, we apply the Fresnel approximation using a second-order Taylor expansion~\cite{wang2023near}, which yields
\begin{equation}
\label{eq:eq-13}
\begin{aligned}
r_n &\approx r_0\left( {1 + \dfrac{{{\delta _n^2}{d^2}}}{{2{r_0^2}}} - \dfrac{{{\delta _n}d}}{r_0}\cos \theta  - \dfrac{{{\delta _n^2}{d^2}{{\cos }^2}\theta }}{{2{r_0^2}}}} \right)\\
 &= r_0 - {\delta _n}d \alpha + \dfrac{{{\delta _n^2}{d^2}}}{{2r_0}}{(1 - \alpha^2)},
\end{aligned}
\end{equation}
where $r_0$ denotes the reference distance and $\alpha = \cos \theta$. This approximation retains quadratic phase terms to capture the spherical wavefront curvature. It is valid for targets located in the radiative NF region, defined by $D < r < 2 D^2 / \lambda_c$ (where $D \approx Nd$ is the aperture size), which ensures the higher-order terms in the Taylor expansion are negligible.

\newcounter{TempEqCnt1}
\setcounter{TempEqCnt1}{\value{equation}} 
\setcounter{equation}{13} 
\begin{figure*}[ht]
\begin{equation}
\label{eq:eq-14}
\begin{aligned}
g\left( {{r_{i}},{\theta _{i}},{r_{j}},{\theta _{j}}} \right) &\triangleq g\left( {{r_{i}},{\alpha _{i}},{r_{j}},{\alpha _{j}}} \right) = \frac{1}{{NM}} \left| {\sum\limits_{m = 0}^{M - 1} {\sum\limits_{n = 0}^{N - 1} {{e^{\mathrm{j}{k_m}({r_{i}} - {r_{j}})}}{e^{\mathrm{j}{k_m}{\delta _n}d\left( {{\alpha _{j}} - {\alpha _{i}}} \right)}}{e^{\mathrm{j}{k_m}\frac{{\delta _n^2{d^2}}}{2}\left( {\frac{{1 - \alpha _{i}^2}}{{{r_{i}}}} - \frac{{1 - \alpha _{j}^2}}{{{r_{j}}}}} \right)}}} } } \right|\\
 &\approx \frac{1}{{NM}} \underbrace {\left| {\sum\limits_{m = 0}^{M - 1} {e^{\mathrm{j}{k_m}({r_{i}} - {r_{j}})}} } \right|}_{{\text{Distance Correlation}}} \cdot \underbrace {\left| {\sum\limits_{n' =  - N'}^{N'} {{e^{\mathrm{j}{k_c}n'd\left( {{\alpha _{j}} - {\alpha _{i}}} \right)}}{e^{\mathrm{j}{k_c}{{\left( {n'} \right)}^2}{d^2}\frac{1}{2}\left( {\frac{{1 - \alpha _{i}^2}}{{{r_{i}}}} - \frac{{1 - \alpha _{j}^2}}{{{r_{j}}}}} \right)}}} } \right|}_{{\text{Angular Correlation}}}.
\end{aligned}
\end{equation}
\hrulefill \vspace{-2pt}
\end{figure*}
\setcounter{equation}{\value{TempEqCnt1}} 
\addtocounter{equation}{1} 

To facilitate the analytical derivation of the sampling grid, we temporarily adopt the NB assumption $f_c \gg B$.
Although the considered system is a broadband system, this assumption implies that the spatial phase variations across the bandwidth are negligible specifically for the purpose of coherence structure analysis. This simplification allows the mutual coherence to be approximately factorized into the product of a distance-dependent frequency term and an angle-distance coupled spatial term as expressed in Eq.~\eqref{eq:eq-14}:
\setcounter{equation}{14}
\begin{equation}
\label{eq:eq-15}
\begin{aligned}
S_{\text{freq}} &= \frac{1}{M}\left|\sum_{m=0}^{M-1} e^{\mathrm{j} k_m (r_{i} - r_{j})}\right|,\\
S_{\text{space}} \! &= \!\frac{1}{N}\!\left|\!\sum_{n'=-N'}^{N'}
e^{\mathrm{j} k_c\, n'd(\alpha_{j} - \alpha_{i})}\,
e^{\mathrm{j} k_c\, (n')^2 d^2\!\left(\frac{1 - \alpha_{i}^2}{2r_{i}} - \frac{1 - \alpha_{j}^2}{2r_{j}}\right)}\right|,
\end{aligned}
\end{equation}where $S_{\text{freq}}$ characterizes the frequency-domain correlation depending on the bandwidth and distance difference, while $S_{\text{space}}$ represents the spatial aperture correlation determined by both angular and distance parameters. Minimizing the total coherence $g(\cdot) \approx S_{\text{freq}} \cdot S_{\text{space}}$ requires at least one of these factors to be negligible. Recognizing their distinct dependencies, we adopt a decoupled design strategy:
\begin{enumerate}
    \item \textbf{Inter-Distance Orthogonalization via} $S_{\text{freq}}$: Since $S_{\text{freq}}$ depends essentially on the distance difference $\Delta r$, it provides robust orthogonality for targets in different distance bins, regardless of their angular separation.
    \item \textbf{Intra-Distance discrimination via} $S_{\text{space}}$: When targets share the same distance bin (collapsing $S_{\text{freq}}$ to unity), orthogonality must be recovered solely through $S_{\text{space}}$.
\end{enumerate}

First, evaluating the summation in $S_{\text{freq}}$ yields
\begin{equation}
\label{eq:eq-16}
\begin{aligned}
S_{\text{freq}} = \frac{1}{M} \left| \frac{\sin\left( \frac{M \Delta k (r_{i} - r_{j})}{2} \right)}{\sin\left( \frac{\Delta k (r_{i} - r_{j})}{2} \right)} \right|,
\end{aligned}
\end{equation}
where $\Delta k = 2\pi \Delta f / c$ denotes the wave number increment. The nulls of Eq.~\eqref{eq:eq-16} establish the fundamental frequency-domain orthogonality condition as follows:
\begin{equation}
\label{eq:eq-17}
\Delta r = r_{i} - r_{j} = \frac{2\pi p}{M \Delta k} = p \cdot \frac{c}{B}, \quad p \in \mathbb{Z} \setminus \{0\}.
\end{equation}
Specifically, the distance resolution determined by the first null is $\Delta r = \frac{2\pi}{M\,\Delta k} = \frac{c}{M\Delta f} = \frac{c}{B}$, which links the distance sampling step directly to the system bandwidth. This confirms that as long as the dictionary grid points are separated by integer multiples of the distance resolution $c/B$, inter-distance coherence is suppressed. The remaining challenge is to design the grid to handle the intra-distance (same $r$) and intra-angle (same $\alpha$, proximate $r$) ambiguities.

\vspace{-4pt}
\subsection{Angular Sampling Design}
When two dictionary atoms share the same distance index ($r_{i} = r_{j} = r$), the frequency term $S_{\text{freq}}$ becomes unity, and distance discrimination fails. Under this condition, the orthogonality must be ensured by the spatial term $S_{\text{space}}$.

\subsubsection{Design of Spatial Sampling Grid}
The angular sampling design focuses on resolving targets when distance discrimination fails. In Eq.~\eqref{eq:eq-15}, the spatial coherence $S_{\text{space}}$ contains both a linear phase term depending on angular difference and a quadratic phase term. Ideally, the quadratic phase is eliminated when the atom pair satisfies the wavefront-curvature matching condition $\frac{1-\alpha_{i}^2}{2r_{i}} = \frac{1-\alpha_{j}^2}{2r_{j}}$. In the intra-distance scenario ($r_{i} = r_{j}$), the quadratic phase difference is proportional to $(\alpha_{i}^2 - \alpha_{j}^2)/r$. Moreover, considering the typical operational scenario where the target distance $r$ is larger than the array aperture $D$ (i.e., $D < r$), the contribution of this quadratic residual scales by a factor proportional to the aperture-to-distance ratio $D/r$ relative to the linear phase term. Consequently, perfectly matching the curvature is secondary to the dominant linear phase mismatch, allowing $S_{\text{space}}$ to be well-approximated by the standard linear array factor
\begin{equation}
\label{eq:eq-18}
\begin{aligned}
S_{\text{space}} &\approx \frac{1}{N} \left| \frac{\sin \left( \frac{N \pi}{2} (\alpha_{j} - \alpha_{i}) \right)}{\sin \left( \frac{\pi}{2} (\alpha_{j} - \alpha_{i}) \right)} \right|,
\end{aligned}
\end{equation}
where the last equality assumes half-wavelength spacing $d=\lambda_c/2$, implying $k_c d=\pi$. Under these conditions, the column coherence function depends purely on the angular difference. Considering the field of view $60^\circ  < \theta  < 120^\circ $ as shown in Fig.~\ref{fig:fig-1}, the corresponding directional cosine $\alpha = \cos \theta$ is constrained to the interval $-\frac{1}{2} < \alpha < \frac{1}{2}$. The nulls of this function provide the angular orthogonality condition
\begin{equation}
\label{eq:eq-19}
\alpha_{j} - \alpha_{i} = \frac{2p}{N}, \quad p \in \mathbb{Z} \setminus \{0\}, \quad |p| \le \frac{N-1}{2}.
\end{equation}

\subsubsection{Angular Sampling Criterion}
Combining this with the bounded field of view $-\frac{1}{2} < \alpha < \frac{1}{2}$, we propose a uniform grid in the directional cosine domain
\begin{equation}
\label{eq:eq-20}
\alpha_{n'} = -\frac{1}{2} + \frac{2n'+1}{N}, \quad n'=0,\ldots,\left\lfloor \frac{N}{2}\right\rfloor -1.
\end{equation}
This grid guarantees that for any same distance targets, their angular difference aligns with the zeros of the spatial coherence function.

\vspace{-8pt}
\subsection{Distance Sampling Design}
While angular sampling relies on linear phase variations, distance sampling in the NF regime necessitates a hybrid approach that exploits both the bandwidth-dependent delay and the wavefront curvature.

\subsubsection{Fresnel-Based Spatial Coherence Analysis}
To isolate the distance-dependent spatial information, we examine the column coherence for two atoms sharing the same angle ($\alpha_{i} = \alpha_{j} = \alpha$) but reside at different distances. Under this condition, the linear phase term in $S_{\text{space}}$ vanishes. The remaining phase term is purely quadratic, simplifying $S_{\text{space}}$ to a function of the inverse-distance difference. This quadratic summation lacks a simple closed-form expression. However, for a sufficiently large number of array elements $N$, the summation can be accurately approximated by a definite integral
\begin{equation}
\label{eq:eq-21}
\begin{aligned}
S_{\text{space}} \approx \left| {F\left( x \right)} \right| &= \frac{1}{N} \left|\sum\limits_{n' = - N'}^{N'} {e^{\mathrm{j} \frac{\pi (n')^2 d^2 (1 - \alpha^2)}{\lambda_c} \left(\frac{1}{r_{i}} - \frac{1}{r_{j}}\right)}} \right|\\
&\approx \frac{1}{N} \left| \int_{-N'}^{N'} e^{\mathrm{j}\pi (n')^2 x} \,dn' \right|,
\end{aligned}
\end{equation}
where $x \triangleq \frac{ d^2 (1 - \alpha^2)}{\lambda_c} (\frac{1}{r_{i}} - \frac{1}{r_{j}})$ is the dimensionless variable characterizing the distance disparity.

By leveraging the symmetry of the integrand and applying the variable substitution $\kappa = n'\sqrt{2|x|}$, the integral is formulated in terms of Fresnel integrals. For the case $x > 0$,
\begin{equation}
\label{eq:eq-22}
\begin{aligned}
F(x)\! \!&\approx\!\! \frac{\sqrt2}{N\sqrt{x}}\!\! \int_0^{\sqrt{2x} N'}\!\! \!\!e^{ \mathrm{j}\frac{\pi}{2}\kappa^2 } \!\!\,d\kappa\! =\! \frac{\sqrt2}{N\sqrt{x}} \! \left[ U(\zeta) \!+\! \mathrm{j}V(\zeta) \right],
\end{aligned}
\end{equation}
where the integration upper limit is $\zeta = \sqrt{2x} N'$. The terms $U(\zeta)$ and $V(\zeta)$ denote the Fresnel integrals, defined as
\begin{equation}
\label{eq:eq-23}
\begin{aligned}
U(\zeta) &= \int_0^{\zeta} \cos \left( \frac{\pi}{2}\kappa^2 \right) \,d\kappa, \\
V(\zeta) &= \int_0^{\zeta} \sin \left( \frac{\pi}{2}\kappa^2 \right) \,d\kappa.
\end{aligned}
\end{equation}
These functions characterize the oscillatory nature of ${F\left( x \right)}$ and facilitate a computationally efficient approximation of its evaluation~\cite{cui2022channel}. Consider another scenario where $r_{i} > r_{j}$, and let $x' = -x > 0$. Through a symmetric derivation, we obtain
\begin{equation}
\label{eq:eq-24}
\begin{aligned}
F(x) &= \frac{2}{N\sqrt{2x'}} \int_0^{\sqrt{2x'} N'} \exp \left( -\mathrm{j}\frac{\pi}{2}\kappa^2 \right) \,d\kappa \\
&= \frac{2}{N\sqrt{2x'}} \left[ U(\zeta) - \mathrm{j}V(\zeta) \right],
\end{aligned}
\end{equation}
where $\zeta  = \sqrt {2x'} N'$. This result mirrors $x>0$, with the sign inversion in the exponential leading to a conjugate relationship.

Consequently, combining both cases (Eq.~\eqref{eq:eq-22} and Eq.~\eqref{eq:eq-24}), the magnitude function $|F(x)|$ relies solely on quadratic sums of $U$ and $V$, independent of the sign of $x$
\begin{equation}
\label{eq:eq-25}
\left| F(x) \right| \approx \frac{2}{N\sqrt{2|x|}} \sqrt{U^2(\zeta) + V^2(\zeta)}, \quad  \zeta = N' \sqrt{2|x|}.
\end{equation}
Asymptotically, as $\zeta \to \infty$, both $U(\zeta)$ and $V(\zeta)$ converge to $0.5$. This implies that the coherence decays according to $|F(x)| \sim \zeta^{-1}$, confirming that orthogonality improves as the inverse-distance difference $|1/r_{i} - 1/r_{j}|$ increases.

\subsubsection{Bandwidth Constraint and Grid Alignment}
While the Fresnel-based analysis (Eq.~\eqref{eq:eq-21}--\eqref{eq:eq-25}) reveals how spatial curvature contributes to distance discrimination, the ultimate distance resolution is hard-limited by the system bandwidth.
To ensure global orthogonality across the dictionary, we enforce the frequency-domain condition derived in Eq.~\eqref{eq:eq-17}. This imposes a discrete grid constraint on the distance sampling, i.e., any two distinct distance geometric centers $r_{i}$ and $r_{j}$ must be separated by an integer multiple of the fundamental resolution $c/B$.
Consequently, the continuous search space for distance sampling is discretized into a uniform candidate set $\mathcal{G}_{\text{grid}} = \{r_{\min} + k \cdot c/B \mid k \in \mathbb{Z}_{\ge 0}\}$. Potential distance atoms will be selected from this grid to maximize the spatial orthogonality derived in the previous subsection.

Our proposed distance sampling strategy harmonizes the Fresnel-based orthogonality derived from Eq.~\eqref{eq:eq-25} with the bandwidth-based grid constraint in Eq.~\eqref{eq:eq-17}. We proceed in two steps: First, we impose a coherence threshold $\Delta \in (0,1)$ such that $|F(x)| \le \Delta$. Let $\zeta_\Delta$ be the value satisfying $|F(\zeta_\Delta)| = \Delta$. The inequality implies a lower bound on the inverse-distance difference
\begin{equation}
\label{eq:eq-26}
\left|\frac{1}{r_{i}}-\frac{1}{r_{j}}\right| \ge \frac{1}{G_\Delta\left(1-\alpha^{2}\right)},
\quad \text{where } G_\Delta \approx \frac{D^{2}}{2\zeta_\Delta^{2}\lambda_c}.
\end{equation}
Enforcing equality in Eq.~\eqref{eq:eq-26} for successive distance samples yields a set of \textit{non-uniform distance rings} distributed along each angular ray~\cite{cui2022channel}, given by
\begin{equation}
\label{eq:eq-27}
\tilde{r}_l(\alpha)\;=\;{G_\Delta\!\left(1-\alpha^{2}\right)}/{l},
\qquad
l=l_{\min},\dots,L,
\end{equation}
where the index $l$ scales linearly with the inverse distance. The starting index $l_{\min}$ accounts for the FF boundary, while $L$ corresponds to the NF lower boundary $r_{\min}$.

Second, to satisfy the bandwidth constraint, we align these ideal rings $\{\tilde{r}_l\}$ onto the resolution grid. Adopting a recursive ``outside-in'' approach, let $r'_{l-1}$ be the aligned position of the previous (outer) ring. The current ring is adjusted to
\begin{equation}
\label{eq:eq-28}
r'_l = r'_{l-1} - k_l \Delta r, \quad
\text{where } k_l = \left\lceil \frac{r'_{l-1} - \tilde{r}_l}{\Delta r} \right\rceil \in \mathbb{Z}_{\ge 1},
\end{equation}
with $\Delta r = c/B$. Here, the ceiling operation guarantees that the inter-ring interval is expanded to the nearest integer multiple of the grid resolution. This expansion ensures that the actual physical distance $r'_{l-1} - r'_l$ is strictly no less than the theoretical requirement $r'_{l-1} - \tilde{r}_l$, thereby preserving the orthogonality conditions provided by the Fresnel bounds.

It remains to be shown that this grid alignment does not violate the Fresnel-based inverse-distance lower bound specified in Eq.~\eqref{eq:eq-26}. Let us adopt an "expansion" strategy. Consider the adjacent pair of aligned rings $(r'_l, r'_{l-1})$, where $r'_l < r'_{l-1}$. The alignment process effectively widens the physical gap by moving the inner ring $r_l$ closer to the origin (decreased by $b \ge 0$) and/or the outer ring $r_{l-1}$ further away (increased by $a \ge 0$), i.e., $r_l' = r_l - b$ and $r_{l-1}' = r_{l-1} + a$. The change in the inverse-distance difference is given by
\begin{equation}
\label{eq:eq-29}
\begin{aligned}
\Delta_{\text{inv}} &= \left( {\frac{1}{{r_l - b}} - \frac{1}{{r_{l - 1} + a}}} \right) - \left( {\frac{1}{{{\tilde{r}_l}}} - \frac{1}{{{\tilde{r}_{l - 1}}}}} \right) \\
&= \frac{b}{{r_l}(r_l - b)} + \frac{a}{{r_{l - 1}}({r_{l - 1}} + a)}.
\end{aligned}
\end{equation}
Given that $r_l > b$ and all distance parameters are positive, the term in Eq.~\eqref{eq:eq-29} remains strictly positive. Therefore,
\begin{equation}
\label{eq:eq-30}
\left|\frac{1}{r'_l}-\frac{1}{r'_{l-1}}\right| > \left|\frac{1}{\tilde{r}_l}-\frac{1}{\tilde{r}_{l-1}}\right| \ge\;\frac{1}{G_\Delta(1-\alpha^{2})}.
\end{equation}
This confirms that increasing the distance gap $|r'_{l-1}-r'_l|$ guarantees a larger inverse-distance gap. Thus, the alignment satisfies the Fresnel-imposed orthogonality lower bound.

\begin{table}[!t]
\centering
\caption{Complexity Comparison of Different Dictionary}
\label{tab:complexity}
\renewcommand{\arraystretch}{1.5} 
\resizebox{\columnwidth}{!}{ 
\begin{tabular}{|l|c|c|}
\hline
\textbf{Scheme} & \textbf{Distance Grid Size ($Q_r$)} & \textbf{Dictionary Complexity} \\ \hline
Traditional Uniform Grid & $\mathcal{O}\left( \frac{r_{\max}}{\epsilon} \right)$ & $\mathcal{O}\left(NM \cdot Q_\theta \cdot \frac{r_{\max}}{\epsilon}\right)$ \\ \hline
Pure Fresnel Grid~\cite{cui2022channel} & $Q_{r,\text{dense}} = \mathcal{O}\left(\frac{1}{r_{\min}}\right)$ & $\mathcal{O}\left(NM \cdot Q_\theta Q_{r,\text{dense}}\right)$ \\ \hline
\textbf{Proposed Hybrid Grid} & $\mathcal{O}\left(\min\left(L, \frac{r_{\max}}{\Delta r}\right)\right)$ & $\mathcal{O}\left(NM \cdot Q_\theta Q_{r}\right)$ \\ \hline
\end{tabular}
}\vspace{-4pt}
\end{table}

\subsubsection{Dimensionality Reduction and Complexity Analysis}

The complexity comparison in Table~\ref{tab:complexity} illustrates the efficiency of our hybrid design compared to traditional schemes. For the traditional uniform grid, the distance atom count is determined by a fixed fine step $\epsilon$. For the pure Fresnel grid~\cite{cui2022channel}, the number of distance atoms is denoted as $Q_{r,\text{dense}} \approx L$, where $L$ is the maximum index defined in Eq.~\eqref{eq:eq-27}. As derived, $L \propto 1/r_{\min}$, which implies that the atom density inflates as the target approaches the array, leading to grid intervals that are physically indistinguishable. In contrast, our hybrid grid truncates the sampling cardinality using the system bandwidth resolution. As shown in Table~\ref{tab:complexity}, the distance grid size $Q_r$ is bounded by $\min(L, r_{\max}/\Delta r)$, where $r_{\max}$ represents the FF distance boundary as mentioned above. Since $r_{\max}/\Delta r$ remains constant regardless of how small $r_{\min}$ is, this strategy ensures that the dictionary size $Q_{\theta}Q_r$ remains scalable even in extreme XL-MIMO configurations. This reduction translates directly to the decrease in the computational complexity of the subsequent CS-based sparse recovery, rendering the unified WB-NF sensing algorithm highly viable for practical 6G ISAC implementations.


\begin{algorithm}[t]
\caption{WB-NF Parameter Estimation Algorithm}
\label{alg:wide_nf}
\begin{algorithmic}[1]
\small

\Statex \textbf{1. Angle grid construction (Eq.~\eqref{eq:eq-20}):}
\State $\alpha_n \gets -\frac{1}{2} + \frac{2n + 1}{N}, \quad n = 0, \dotsc, \lfloor N/2 \rfloor - 1$

\Statex \textbf{2. Non-uniform distance rings alignment (Eq.~\eqref{eq:eq-28}):}
\State $\lambda_c \gets c/f_c$; \quad $G_\Delta \gets \frac{(Nd)^2}{2\zeta_\Delta^2 \lambda_c}$; \quad $\Delta r \gets \frac{c}{M \Delta f}$; Initialize $\mathcal{G} \gets \emptyset$

\ForAll{$\alpha_n$}
    \State $r'_{\text{prev}} \gets r_{\max}$ \Comment{Initialize at FF boundary}
    \State $l \gets \lceil G_\Delta(1 - \alpha_n^2) / r_{\max} \rceil$ \Comment{Determine initial index}
    \While{true}
        \State $\tilde{r}_l \gets \frac{G_\Delta(1 - \alpha_n^2)}{l}$ \Comment{Ideal candidate ring}
        \State $k_l \gets \lceil (r'_{\text{prev}} - \tilde{r}_l) / \Delta r \rceil$
        \State $r'_l \gets r'_{\text{prev}} - k_l \Delta r$ \Comment{Quantize to resolution grid}

        \If{$r'_l < r_{\min}$} \textbf{break} \EndIf

        \State Add pair $(\alpha_n, r'_l)$ to set $\mathcal{G}$; $r'_{\text{prev}} \gets r'_l$; \quad $l \gets l + 1$
    \EndWhile
\EndFor

\Statex \textbf{3. Dictionary construction:}
\State $Q_{\theta}Q_r \gets |\mathcal{G}|$; Initialize $\mathbf{B} \in \mathbb{C}^{NM \times Q_{\theta}Q_r}$
\For{$q = 1$ to $Q_{\theta}Q_r$}
    \State Extract $(\alpha_q, r_q)$ from $\mathcal{G}$
    \For{$m = 0$ to $M-1$}
        \State $k_m \gets \frac{2\pi(f_c + m \Delta f)}{c}$
        \State Calculate exact distance $r_{n}(\alpha_q, r_q)$ via Eq.~\eqref{eq:eq-1}
        \State Define $[\mathbf{\tilde a}_m]_n = \frac{1}{\sqrt{NM}} e^{-\mathrm{j} k_m r_{n}(\alpha_q, r_q)}$
        \State $\mathbf{B}_{mN+1:(m+1)N, q} \gets \mathbf{\tilde a}_m$
    \EndFor
\EndFor

\Statex \textbf{4. Sparse recovery and parameter extraction:}
\State $(\sim, \mathcal{S}) \gets \text{SOMP}(\mathbf{Y}, \mathbf{B}, P)$ \Comment{Estimate support set}
\For{each index $s \in \mathcal{S}$}
    \State Retrieve estimated $(\hat{\alpha}_s, \hat{r}_s)$ from $\mathcal{G}$ using index $s$
    \State $\hat{\theta}_p \gets \arccos(\hat{\alpha}_s)$; $\hat{r}_p \gets \hat{r}_s$
    \State Add estimated $(\hat{r}_p, \hat{\theta}_p)$ to output set
\EndFor
\end{algorithmic}
\vspace{-2pt}
\end{algorithm}

\vspace{-2pt}
\subsection{The Proposed Algorithm for WB-NF Parameter Estimation}
The proposed estimation procedure is outlined in Algorithm \ref{alg:wide_nf}, which comprises four distinct phases:

\begin{enumerate}
 \item \textbf{Angle Grid Initialization:}
    First, a uniform angular grid is initialized in the spatial frequency domain according to Eq.~\eqref{eq:eq-20}. This establishes the directional baselines ensuring that the linear phase component of the spatial coherence is minimized for targets at similar distances.

 \item \textbf{Recursive Hybrid Distance Sampling:}
    Second, for each angular ray $\alpha_n$, a set of non-uniform distance rings is generated using a recursive ``outside-in'' strategy. The search initiates from the FF boundary $r_{\max}$ (setting $r'_{\text{prev}} \leftarrow r_{\max}$). In each iteration, the algorithm computes two candidates: 1) The \textit{ideal Fresnel distance} $\tilde{r}_l$, calculated via Eq.~\eqref{eq:eq-27}, which satisfies the inverse-distance orthogonality condition derived from the Fresnel approximation. 2) The \textit{aligned distance} $r'_l$, obtained by quantizing the interval relative to the previous ring $r'_{\text{prev}}$ onto the system's distance resolution $\Delta r$, as defined in Eq.~\eqref{eq:eq-28}.
    This hybrid alignment suppresses both spatial correlation (via $\tilde{r}_l$) and frequency-domain sidelobes (via grid alignment). The recursion proceeds by incrementing the ring index $l$ until the lower boundary $r_{\min}$ is reached.

\item \textbf{Exact Dictionary Construction:}
    Third, the WB overcomplete dictionary $\mathbf{B}$ is assembled. Here, $Q_{\theta}Q_r \leftarrow |\mathcal{G}|$ represents the total cardinality of the generated non-uniform grid points, which determines the column dimension of the dictionary matrix $\mathbf{B}$. A key distinction in this step is the model fidelity: while the sampling grid $\mathcal{G}$ is designed using Fresnel approximations to ensure low coherence, the dictionary atoms are constructed using the exact spherical wavefront model. Specifically, for each grid point $(\alpha_q, r_q)$ in $\mathcal{G}$, the precise array delays are calculated via Eq.~\eqref{eq:eq-1}, and the corresponding joint spatial-frequency steering vectors are normalized to unit norm ($1/\sqrt{NM}$) to form the columns of $\mathbf{B}$.

\item \textbf{Sparse Recovery and Extraction:}
    Finally, the Simultaneous Orthogonal Matching Pursuit (SOMP) algorithm is employed to identify the support set $\mathcal{S}$ from the received signal $\mathbf{Y}$ and dictionary $\mathbf{B}$. The estimated indices are then mapped back to the physical grid parameters to retrieve the target locations $\{(\hat{r}_p, \hat{\theta}_p)\}_{p=1}^P$.
\end{enumerate}

\section{Decoupled Analysis of the WB and NF Effects}
To isolate and quantitatively assess the individual contributions of the NF and WB effects, we introduce two auxiliary benchmark schemes. These schemes are constructed by deliberately disabling one of the two physical phenomena in the signal model, thereby enabling a clean decomposition of their impact on parameter estimation performance. Specifically: 1) NB-NF MUSIC: This scheme considers only the spherical wavefront curvature effect while neglecting the frequency-dependent effect. 2) WB-FF MUSIC: This scheme considers only the WB effect while assuming planar wavefronts.

\subsection{The NB-NF Boundary and MUSIC Method}
\label{sec:IV-A}
\subsubsection{Correlation Function and Dirichlet-Kernel Bound}
\label{sec:IV-A-1}
To examine the impact of WB effects, we define the spatial correlation function as the magnitude of the normalized inner product between the WB-NF steering vector and its NB reference counterpart constructed at the central carrier wavenumber $k_c$.
This correlation metric, which takes values in the interval $[0,1]$, quantifies the frequency-dependent mismatch between the actual WB snapshots and the reference model, thereby capturing the decorrelation induced by the WB effect even when the spatial direction remains fixed. Incorporating the NF distance approximation derived in Eq.~(\ref{eq:eq-13}), the normalized correlation between the WB-NF steering vector and the NB reference steering vector is derived as Eq.~(\ref{eq:eq-32}).
\newcounter{TempEqCnt2}
\setcounter{TempEqCnt2}{\value{equation}}
\setcounter{equation}{30}
\begin{figure*}[ht]
\begin{equation}
\label{eq:eq-32}
\begin{aligned}
   \tilde{g}_1(r_{i}, \theta_{i}) &\triangleq \tilde g_1\left( {{r_{i}},{\alpha _{i}}} \right) =  \bigl| \mathbf{a}^H(r_{i}, \theta_{i}) \mathbf{\tilde a}_{1}(r_{i}, \theta_{i}) \bigr|
   = \frac{1}{NM} \left| \sum_{m=0}^{M-1} \sum_{n=0}^{N-1} e^{\mathrm{j} k_m r_{i,n}} e^{-\mathrm{j} k_c r_{i,n}} \right| \\
   &= \frac{1}{NM} \left| \sum_{m=0}^{M-1} e^{\mathrm{j} m \Delta k r_{i}} \sum_{n'=-N'}^{N'}
      \underbrace{ e^{-\mathrm{j} m \Delta k n' d \alpha_{i} } }_{\text{Linear Phase}}
      \underbrace{ e^{\mathrm{j} m \Delta k \frac{(n')^2 d^2 (1-\alpha_{i}^2)}{2 r_{i}}} }_{\text{Quadratic Phase}} \right|.
\end{aligned}
\end{equation}
\hrulefill \vspace{-2pt}
\end{figure*}
\setcounter{equation}{\value{TempEqCnt2}}

Assuming a fixed spatial direction described by the directional cosine $\alpha = \cos\theta$, we consider $M$ discrete wavenumbers $k_m = k_c + m \Delta k$ for $m = 0, \ldots, M - 1$, where the total wavenumber span is defined as $K_{bw} = M \Delta k$. For each antenna index $n'$, let us define the effective Fresnel distance term $\psi_{n'}(r) = r - n'd\,\alpha + \frac{(n')^2 d^2}{2r}(1-\alpha^2).$ Substituting this into Eq.~(\ref{eq:eq-32}), the summation over the frequency domain (index $m$) typically takes the form of a geometric series. For a generic distance variable $\psi$, this summation is expressed as
\setcounter{equation}{31}
\begin{equation}
\label{eq:eq-33}
\begin{aligned}
\sum_{m=0}^{M-1} e^{\mathrm{j}\,m\Delta k\, \psi} &= e^{\mathrm{j}\frac{(M-1)\Delta k\, \psi}{2}}\,
\frac{\sin\!\left(\frac{M\Delta k\, \psi}{2}\right)}{\sin\!\left(\frac{\Delta k\,\psi}{2}\right)} \\
&\triangleq e^{\mathrm{j}\frac{(M-1)\Delta k\, \psi}{2}}\,
\mathcal{D}_M(\Delta k\, \psi),
\end{aligned}
\end{equation}
where $\gamma = \Delta k\,\psi$, and $\mathcal{D}_M(\gamma) \triangleq \frac{\sin(M \gamma/2)}{\sin(\gamma /2)}$ denotes the Dirichlet kernel (also known as the periodic sinc function). This derivation reveals that aggregating $M$ subcarriers results in an amplitude envelope shaped by the Dirichlet kernel, exhibiting a characteristic mainlobe-sidelobe structure that dictates the distance resolution. By substituting the antenna-specific term $\psi_{n'}(r)$ back into the array correlation expression in Eq.~(\ref{eq:eq-32}), we obtain the closed-form expression for the normalized WB-NF correlation
\begin{equation}
\label{eq:eq-34}
\begin{aligned}
\tilde{g}_1(r_{i}, \alpha_{i})
&= \frac{1}{NM} \left| \sum_{n'=-N'}^{N'} \left( \sum_{m=0}^{M-1} e^{\mathrm{j} m \Delta k \psi_{n'}(r)} \right) \right| \\
&= \frac{1}{NM} \left| \sum_{n'=-N'}^{N'} \mathcal{D}_M \bigl(\Delta k\, \psi_{n'}(r)\bigr) \, e^{ \mathrm{j} \frac{(M - 1)\Delta k}{2} \psi_{n'}(r) } \right|.
\end{aligned}
\end{equation}
From Eq.~(\ref{eq:eq-34}), the magnitude contribution from each antenna element is $\frac{1}{M} \left| \mathcal{D}_M(\Delta k \psi _{n'}(r)) \right|$, whose mainlobe is confined by the first nonzero roots of $\sin(M \gamma/2)$ at $\gamma = \pm 2\pi/M$. In contrast, the periodic peaks (grating lobes, where the denominator also vanishes) occur at integer multiples of $2\pi$. Since $2\pi/M \ll 2\pi$ for any typical number of subcarriers $M \ge 2$, focusing on the mainlobe region is sufficient for local analysis. Consequently, to ensure that the correlation energy remains concentrated within the mainlobe of the Dirichlet kernel, the following sufficient condition must be satisfied, given by
\begin{equation}
\label{eq:eq-35}
\max_{n'\in[-N',\,N']} \bigl| \Delta k\,\psi_{n'}(r) \bigr| \le \frac{2\pi}{M}.
\end{equation}

It should be noted that Eq.~(\ref{eq:eq-35}) is a sufficient but not necessary condition. More specifically, it guarantees that each element remains within the mainlobe, but does not ensure that the overall correlation is close to unity, since the amplitude near the mainlobe edge is already significantly reduced. For a given target angle $\theta_0$, we formally define the maximum threshold distance of the NB-NF region, denoted as $r_{\mathrm{NB-NF}}$, as the solution to the following constrained optimization problem~\cite{10220205}
\begin{subequations}
\label{eq:eq-35ab}
\begin{align}
r_{\mathrm{NB-NF}} =& \arg \max_{r} \quad r, \label{eq:eq-35a} \\
\text{s.t.} \quad & \tilde{g}_1(r, \theta_0) \ge \rho_0, \label{eq:eq-35b}
\end{align}
\end{subequations}
where $\rho_0$ represents a predefined reliability threshold (e.g., $\rho_0 = 0.9$) that ensures the multi-carrier spatial snapshots do not severely decorrelate from the reference NB manifold. By formulating it as Eq.~\eqref{eq:eq-35ab}, $r_{\mathrm{NB-NF}}$ establishes a critical upper boundary. While its exact analytical solution depends on the complex interplay of bandwidth, aperture, and target angle, its quantitative behavior will be rigorously evaluated via simulation in Section~\ref{sec:V-B}.

\subsubsection{The NB-NF MUSIC Algorithm}
Motivated by the definition of $r_{\mathrm{NB-NF}}$ above, we propose an approximation algorithm that leverages the NB-NF model to estimate the location parameters $(r_p, \theta_p)$ from the WB observation. This approach is theoretically grounded in the analysis in Section~\ref{sec:IV-A-1}, which suggests that within the distance $r \le r_{\mathrm{NB-NF}}$, the frequency-dependent phase variations are negligible.

Based on the signal model in Section~\ref{sec:sig}, the aggregated received signal matrix $\mathbf{Y} \in \mathbb{C}^{NM \times K}$ collects the observations over $K$ OFDM symbols collects the observations from all $M$ subcarriers and $N$ antennas. It can be expressed as
\begin{equation}
\label{eq:eq-36}
\mathbf{Y} = {{\mathbf{A}}}\left( \bm {r,\theta } \right) \mathbf{\Gamma} \mathbf{S} + \mathbf{W},
\end{equation}
where $\mathbf{\Gamma} = \mathrm{diag}(\bm{\beta}) \in \mathbb{C}^{P \times P}$ contains the complex path gains, $\mathbf{S} \in \mathbb{C}^{P \times K}$ contains the transmitted symbols, and $\mathbf{W}$ is the additive noise. The matrix ${{\mathbf{A}}}\left( \bm {r,\theta } \right) \in \mathbb{C}^{NM \times P}$ represents the exact WB-NF steering matrix. Its $p$-th column, corresponding to the $p$-th target, stacks the steering vectors across $M$ subcarriers.

To implement the NB approximation, we reorganize the data structure to exploit frequency diversity as multiple snapshots. Let $\mathbf{Y}_m \in \mathbb{C}^{N \times K}$ denote the sub-matrix extracted from $\mathbf{Y}$ corresponding to the $m$-th subcarrier index. The reduced-dimension spatial covariance matrix is then computed by
\begin{equation}
\label{eq:eq-37}
\mathbf{R}_{nb} = \frac{1}{MK} \sum_{m=0}^{M-1} \mathbf{Y}_m \mathbf{Y}_m^H \in \mathbb{C}^{N \times N}.
\end{equation}
Performing the eigendecomposition on $\mathbf{R}$ yields
\begin{equation}
\label{eq:eq-38}
\mathbf{R}_{nb} = \mathbf{U}_s \mathbf{\Sigma}_s \mathbf{U}_s^H + \mathbf{U}_n \mathbf{\Sigma}_n \mathbf{U}_n^H,
\end{equation}
where $\mathbf{U}_n \in \mathbb{C}^{N \times (N-P)}$ represents the noise subspace eigenvectors corresponding to the $N-P$ smallest eigenvalues.

The key to this algorithm is the construction of the approximate scanning vector. Under the NB assumption justified by the correlation analysis above, we assume the frequency-dependent spatial variation is negligible. Consequently, the search vector $\mathbf{a}_{\mathrm{nb}}(r, \theta) \in \mathbb{C}^{N \times 1}$ is simply the standard NF steering vector at the center frequency $f_c$
\begin{equation}
\label{eq:eq-39}
\mathbf{a}_{\mathrm{nb}}(r, \theta) = \mathbf{a}_0(r, \theta),
\end{equation}
where $\mathbf{a}_0(r, \theta) \in \mathbb{C}^{N \times 1}$ is defined in Eq.~(\ref{eq:eq-6}).

Finally, the 2D angle-distance spectrum is given by
\begin{equation}
\label{eq:eq-40}
S_{\mathrm{nb}}(r, \theta) = \frac{1}{\mathbf{a}_{\mathrm{nb}}^H(r, \theta) \mathbf{U}_n \mathbf{U}_n^H \mathbf{a}_{\mathrm{nb}}(r, \theta)}.
\end{equation}
This strategy effectively reduces the computational complexity of manifold generation while maintaining high distance sensing accuracy, provided that the target is located within the valid region $r \le r_{\mathrm{NB-NF}}$ where the WB effect is minimal.

\newcounter{TempEqCnt3}
\setcounter{TempEqCnt3}{\value{equation}}
\setcounter{equation}{40}
\begin{figure*}[ht]
\begin{equation}
\label{eq:eq-41}
\begin{aligned}
   \tilde g_2\left( {{r_{i}},{\theta _{i}}} \right) \triangleq \tilde g_2\left( {{r_{i}},{\alpha _{i}}} \right) &= \left| {{\mathbf{ a}^H}{{({r_{i}},{\theta_{i}})}}{{\mathbf{\tilde a}_{wb}}}({r_{i}},{\theta_{i}})} \right| = \frac{1}{{ N M}} \left| {\sum\limits_{m = 0}^{M - 1} {\sum\limits_{n = 0}^{N - 1} {{e^{\mathrm{j}{k_m} {r_{i,n}}} {e^{-\mathrm{j}{k_m} \tilde r_{i,n}}}} } } } \right|\\
   &= \frac{1}{{N M}} \left| \sum\limits_{n' = -N'}^{N'} e^{\mathrm{j} k_c \frac{(n'd)^2 (1-\alpha_{i}^2)}{2r_{i}}} \left( \sum\limits_{m = 0}^{M - 1} e^{\mathrm{j} m \Delta k \frac{(n'd)^2 (1-\alpha_{i}^2)}{2r_{i}}} \right) \right|.
\end{aligned}
\end{equation}
\hrulefill  \vspace{-4pt}
\end{figure*}

\subsection{The WB-FF Boundary and MUSIC Method}
\subsubsection{Correlation Function and Fresnel-Phase Bound}
To quantify the mismatch purely caused by the wavefront curvature in the NF, we define a normalized inner product—i.e., a correlation function—between a multi-carrier, NF steering vector and a multi-carrier, FF (plane wave) reference vector constructed over the same set of wavenumbers $\{k_m\}_{m=0}^{M-1}$. Unlike the NF effect-only case in Section~\ref{sec:IV-A}, where the reference vector is NB (i.e., constructed at the carrier wavenumber $k_c$), both vectors here are WB. The only difference lies in the retention of the quadratic phase term in the NF model, while the FF model neglects it. This correlation function takes values in $[0,1]$ and isolates the decorrelation solely due to spherical wavefront curvature, with the spatial direction fixed.\vspace{-4pt}

Applying the NF distance approximation from Eq.~(\ref{eq:eq-13}), the resulting normalized correlation is shown in Eq.~(\ref{eq:eq-41}). For a fixed direction $\alpha=\cos\theta$, denote the physical array position by $\chi=n'd$, and define the curvature factor $\eta \triangleq \frac{1-\alpha^2}{2r_{i}}$. Then the residual distance offset between the NF and FF models reduces to a purely quadratic term
\setcounter{equation}{41}
\begin{equation}
\label{eq:eq-42}
{r_{i, n}} -\tilde r_{i, n} = \chi^2 \eta,
\end{equation}
so that the inner summation over subcarriers in Eq.~(\ref{eq:eq-41}) decouples into a product of a frequency-independent curvature phase and a frequency-dependent summation, and we obtain
\begin{equation}
\label{eq:eq-43}
\begin{array}{c}
\begin{aligned}
\sum_{m=0}^{M-1} e^{\mathrm{j}\,m\Delta k\,(\chi^2\eta)}
&= e^{\mathrm{j}\frac{(M-1)\Delta k\,\chi^2\eta}{2}}\,
\frac{\sin\!\big(\tfrac{M\Delta k\,\chi^2\eta}{2}\big)}{\sin\!\big(\tfrac{\Delta k\,\chi^2\eta}{2}\big)}\\
&\ \triangleq\ e^{\mathrm{j}\frac{(M-1)\Delta k\,\chi^2\eta}{2}}\,
\mathcal D_M\!\big(\Delta k\,\chi^2\eta\big),
\end{aligned}
\end{array}
\end{equation}
where $\gamma' = \Delta k\,\chi^2\eta$, and $\mathcal{D}_M(\gamma') \triangleq \frac{\sin(M \gamma'/2)}{\sin(\gamma' /2)}$ denotes the Dirichlet kernel. Substituting \eqref{eq:eq-43} into Eq.~(\ref{eq:eq-41}) yields the array-level normalized correlation
\begin{equation}
\label{eq:eq-44}
\begin{array}{c}
\begin{aligned}
&\tilde g_2\left( {{r_{i}},{\alpha _{i}}} \right)\!=\!\frac{1}{NM}\!\left|\!\sum_{n'=-N'}^{N'}\!\!\!\mathcal D_M\!\big(\Delta k \chi^2\eta\big)
e^{\mathrm{j}\frac{(M-1)\Delta k\,\chi^2\eta}{2}} e^{\mathrm{j} k_c\,\chi^2\eta } \right|.
\end{aligned}
\end{array}
\end{equation}
The factor $\exp\{\mathrm{j}\frac{(M-1)\Delta k\,\chi^2\eta}{2}\}$ is a pure phase and does not affect the magnitude. Hence each array element frequency-domain magnitude contribution is $\frac{1}{M}\,\Big|\mathcal D_M\!\big(\Delta k\,\chi^2\eta\big)\Big|$. Similar to Eq.~(\ref{eq:eq-34}), the mainlobe of $\mathcal D_M(x)$ is confined by the first nonzero roots of $\sin(M \gamma'/2)$ at $\gamma'=\pm 2\pi/M$. In contrast, the periodic peaks (grating lobes, where the denominator also vanishes) occur at integer multiples of $2\pi$. Since $2\pi/M \ll 2\pi$ for any typical number of subcarriers $M \ge 2$, focusing on the mainlobe region is sufficient for analyzing the decorrelation effect. Consequently, to ensure that the WB response for every array element remains concentrated within the high-correlation mainlobe of the Dirichlet kernel, the following sufficient condition must be satisfied
\begin{equation}
\label{eq:eq-45}
\max_{\,|n'|\le N'}\;|\Delta k\,\chi^2\eta|\;=\;\Delta k\,\eta\,(N'd)^2\;\le\;\frac{2\pi}{M}.
\end{equation}
Equivalently, in terms of the distance $r$, we have
\begin{equation}
\label{eq:eq-46}
r\;\ge\;\frac{M\,\Delta k\,(1-\alpha^2)\,(N'd)^2}{4\pi}.
\end{equation}
In contrast to the NF effect-only case, Eq.~(\ref{eq:eq-44}) also contains the carrier-wavenumber Fresnel phase $\exp\{\mathrm{j} k_c\chi^2\eta\}$ across the aperture, which can induce additional spatial cancellation. To avoid strong spatial defocusing, one may further impose the Fresnel-type sufficient bound
\begin{equation}
\label{eq:eq-47}
{\ \max_{\,|n'|\le N'}\;|k_c\,\chi^2\eta|\;=\;k_c\,\eta\,(N'd)^2\;\le\;\pi\ },
\end{equation}which translates, using $k_c=2\pi/\lambda_c$ and $D=2N'd$, we have
\begin{equation}
\label{eq:eq-48}
\ r\;\ge\; \frac{(1-\alpha^2)\,D^2}  {{4\,\lambda_c}\ }.
\end{equation}

The bounds in Eq.~\eqref{eq:eq-46} and Eq.~\eqref{eq:eq-48} are sufficient (not necessary) and jointly ensure that curvature-induced decorrelation remains limited both in frequency (via $\mathcal D_M$) and across the aperture (via the Fresnel phase at $k_c$). Similarly, we define the minimum threshold distance, denoted as $r_{\mathrm{WB-FF}}$. For a given target angle $\theta_0$, $r_{\mathrm{WB-FF}}$ is formulated as the solution to the following constrained optimization problem~\cite{10220205}
\begin{subequations}
\label{eq:eq-48ab}
\begin{align}
r_{\mathrm{WB-FF}} =& \arg \min_{r} \quad r, \label{eq:eq-48a} \\
\text{s.t.} \quad & \tilde{g}_2(r, \theta_0) \ge \rho_1, \label{eq:eq-48b}
\end{align}
\end{subequations}
where $\rho_1$ is a predefined reliability threshold (e.g., $\rho_1 = 0.9$). Physically, Eq.~\eqref{eq:eq-48ab} determines the critical distance boundary below which the uncompensated wavefront curvature induces severe decorrelation ($\tilde{g}_2 < \rho_1 $). Similar to the NB-NF boundary, its quantitative behavior and specific numerical limits will be rigorously evaluated via simulation in Section~\ref{sec:V-C}.

\subsubsection{The WB-FF MUSIC Algorithm}
Alternatively, to address the frequency-dependent phase variations while neglecting the NF curvature, we consider the WB-FF approximation. This approach corresponds to the planar reference model analyzed in Eq.~(\ref{eq:eq-41}). Based on this approximation, the received signal model is matched against a WB planar steering matrix, denoted as $\tilde{\mathbf{A}}$. The signal subspace equation is formulated as
\begin{equation}
\label{eq:eq-49}
\mathbf{\tilde Y}  = \tilde{\mathbf{A}}\left( {\bm{r, \theta}} \right){\mathbf{\Gamma}}{{\mathbf{S}}} + {{\mathbf{W}}},
\end{equation}where ${\mathbf{\Gamma}} = \mathrm{diag}(\bm{\beta})$ denotes the path gains, and $\tilde{\mathbf{A}}\left( {\bm{r, \theta}} \right) = \left[ \tilde{\mathbf{a}}_{wb}({r_1},{\theta _1}), \ldots , \tilde{\mathbf{a}}_{wb}({r_P},{\theta _P}) \right]$ denotes the WB steering matrix. Distinct from the NF model in Eq.~(\ref{eq:eq-6}), the planar steering vector $\tilde{\mathbf{a}}_{wb}$ is constructed using the linear phase approximation
\begin{equation}
\label{eq:eq-50}
\tilde{\mathbf{a}}_{wb}\left( {{r_p},{\theta _p}} \right) = {\left[ {\begin{array}{*{20}{c}}
{\tilde{\mathbf{a}}_0^T({r_p},{\theta _p}),}&{ \ldots ,}&{\tilde{\mathbf{a}}_{M - 1}^T({r_p},{\theta _p})}
\end{array}} \right]^T},
\end{equation}
where $\tilde{\mathbf{a}}_m({r_p},{\theta _p}) \in {\mathbb{C}^{N \times 1}}$ represents the FF response vector at subcarrier $m$, defined as
\begin{equation}
\label{eq:eq-51}
\tilde{\mathbf{a}}_m({r_p},{\theta _p}) = {\left[ {\begin{array}{*{20}{c}}
{{e^{ - \mathrm{j}{k_m}\tilde r_{p,0}}},}&{ \ldots ,}&{{e^{ - \mathrm{j}{k_m}\tilde r_{p,{N - 1}}}}}
\end{array}} \right]^T}.
\end{equation}Here, the effective planar distance is $\tilde r_{p,n} = {r_p} - {\delta _n}d\cos{\theta _p}$. Note that unlike the true NF model, the quadratic curvature term is absent in $\tilde r_{p,n}$. To fully preserve the frequency-dependent effect, the covariance matrix cannot be spatially averaged as in the previous NB-NF case. Instead, the full-dimension spatio-temporal covariance matrix $\mathbf{R}_{wb} = \frac{1}{K}\mathbf{\tilde Y}\mathbf{\tilde Y}^H \in \mathbb{C}^{NM \times NM}$ must be constructed. Performing eigendecomposition on $\mathbf{R}_{wb}$ yields the noise subspace $\tilde{\mathbf{U}}_{n} \in \mathbb{C}^{NM \times (NM-P)}$. By utilizing the orthogonality between the full-dimension signal and noise subspaces, the 2D pseudo-spectrum is formulated as
\begin{equation}
\label{eq:eq-52}
{S_{wb}}({r},{\theta}) = \frac{1}{{\tilde{\mathbf{a}}_{wb}^H{{({r},{\theta})}}{\tilde{\mathbf{U}}_{n}} \tilde{\mathbf{U}}_{n}^H{{\tilde{\mathbf{a}}}_{wb}}({r},{\theta})}}.
\end{equation}

By searching for the peaks of ${S_{wb}}({r},{\theta})$, the joint distance and angle estimation is performed. However, distinct from the proposed NF method, this approach assumes a linear spatial phase distribution, effectively discarding the aperture-domain focusing gain associated with the wavefront curvature. Thus, this approach functionally realizes ranging via the WB delay signature and angle estimation via the planar array response, albeit suffering from defocusing errors if the target strictly resides in the NF curvature region $r < r_{\mathrm{WB-FF}}$.

Table~\ref{tab:regime_summary} summarizes the key characteristics of the three sensing regimes. As shown, the simplified NB-NF and WB-FF models are only valid within their specific distance boundaries ($r \le r_{\mathrm{NB-NF}}$ and $r \ge r_{\mathrm{WB-FF}}$). However, in the transition zone ($r_{\mathrm{NB-NF}} < r < r_{\mathrm{WB-FF}}$), neither approximation is reliable. In this region, relying on only NF or WB delay leads to large estimation errors. Therefore, the unified WB-NF model and the proposed CS method are necessary to bridge this gap, ensuring accurate sensing across the entire range.

\begin{table*}[t]
\centering
\caption{Summary of Sensing Regimes, Physical Assumptions, and Valid Boundaries}
\label{tab:regime_summary}
\renewcommand{\arraystretch}{1.0}
\begin{tabular}{l c l l l l}
\toprule
\textbf{Sensing Regime} & \textbf{Valid Range} & \textbf{Wavefront} & \textbf{Bandwidth} & \textbf{Dominant Mechanism} & \textbf{Algorithm} \\
\midrule
\textbf{NB-NF} Model & $r \le r_{\mathrm{NB-NF}}$ & Spherical & NB & Spherical curvature & NB-NF MUSIC \\
\textbf{WB-FF} Model & $r \ge r_{\mathrm{WB-FF}}$ & Planar & WB & Time-delay & WB-FF MUSIC \\
\textbf{WB-NF} (Unified) & $r_{\mathrm{NB-NF}} < r < r_{\mathrm{WB-FF}}$\textsuperscript & Spherical & WB & Joint spherical \& delay & Proposed CS Method\\
\bottomrule
\end{tabular}\vspace{-5pt}
\end{table*}

\section{Numerical Results}
In this section, we present numerical results to evaluate the performance of the proposed WB-NF parameter estimation algorithm. To demonstrate the performance of the unified model, we compare the proposed method against two benchmark schemes: NB-NF MUSIC and WB-FF MUSIC. Unless otherwise specified, the simulation setup employs a ULA with $N = 127$ elements and an array aperture of $D = 0.68$~m. The system operates at a carrier frequency of $f_c = 28$~GHz with a bandwidth of $B = 122.88$~MHz ($M = 256$ subcarriers). Consequently, the Rayleigh distance is derived as $R_r = 2D^2/\lambda_c \approx 86.4$~m. The signal-to-noise ratio (SNR) is set to $0$~dB by default.

\subsection{Performance Analysis of Sparse Matrix Representation}
\subsubsection{Analysis of Spatial Coherence and Parameter Selection}
\begin{figure}
\begin{center}
\includegraphics[width=0.70\linewidth]{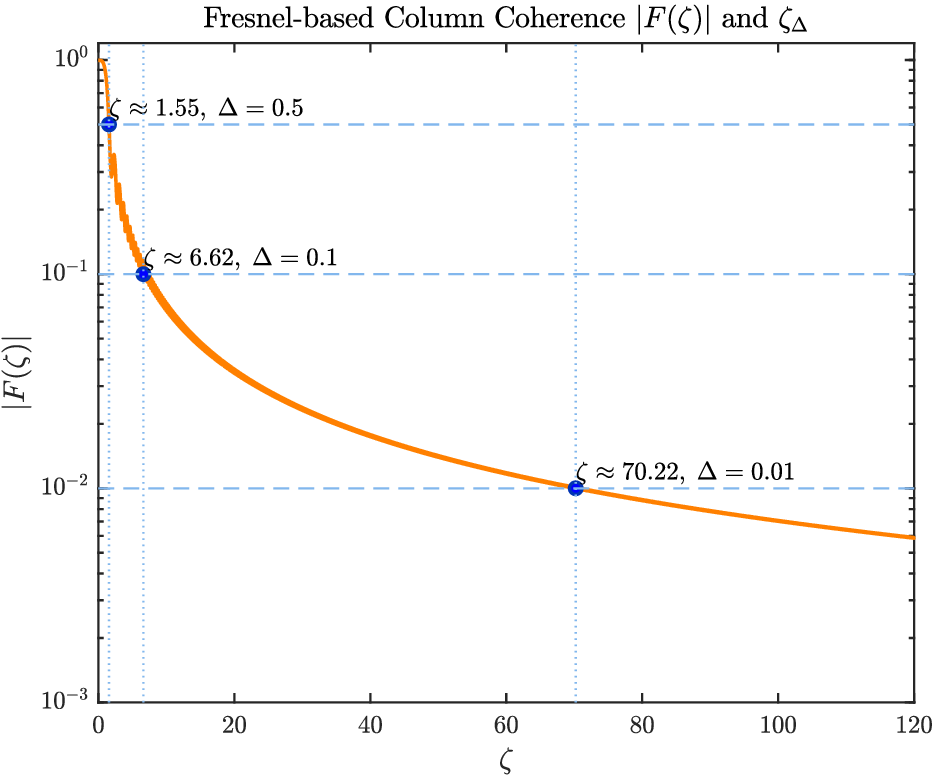}
\end{center}
\caption{Coherence function magnitude $|F(\zeta)|$ as a function of the normalized distance $\zeta$.}
\label{fig:fig-2}\vspace{-8pt}
\end{figure}

To validate the theoretical basis of the proposed dictionary construction, Fig.~\ref{fig:fig-2} illustrates the magnitude of the Fresnel-based column coherence $|F(\zeta)|$ as derived in Eq.~\eqref{eq:eq-25}. The numerical results demonstrate an asymptotic decay behavior proportional to $1/\zeta$. Crucially, the figure reveals the trade-off between dictionary conditioning and grid density involved in selecting the threshold $\Delta$: 1) A strict threshold (e.g., $\Delta = 0.01$) requires a large normalized distance $\zeta \approx 70.22$. This would result in sparse distance rings and increase the risk of off-grid errors. 2)A relaxed threshold (e.g., $\Delta = 0.5$, $\zeta \approx 1.55$) yields dense sampling but results in high mutual coherence, degrading the sparse recovery performance.
Therefore, we adopt $\Delta = 0.1$ ($\zeta_\Delta \approx 6.62$) for the subsequent simulations. This choice offers a favorable balance, ensuring sufficient low-coherence to satisfy the RIP while maintaining adequate sampling resolution for precise localization.

\subsubsection{Validation of the Proposed WB-NF Dictionary Design}
\begin{figure}[!t]
    \centering
    \begin{subfigure}[b]{0.50\columnwidth} 
        \centering
        \includegraphics[width=\linewidth]{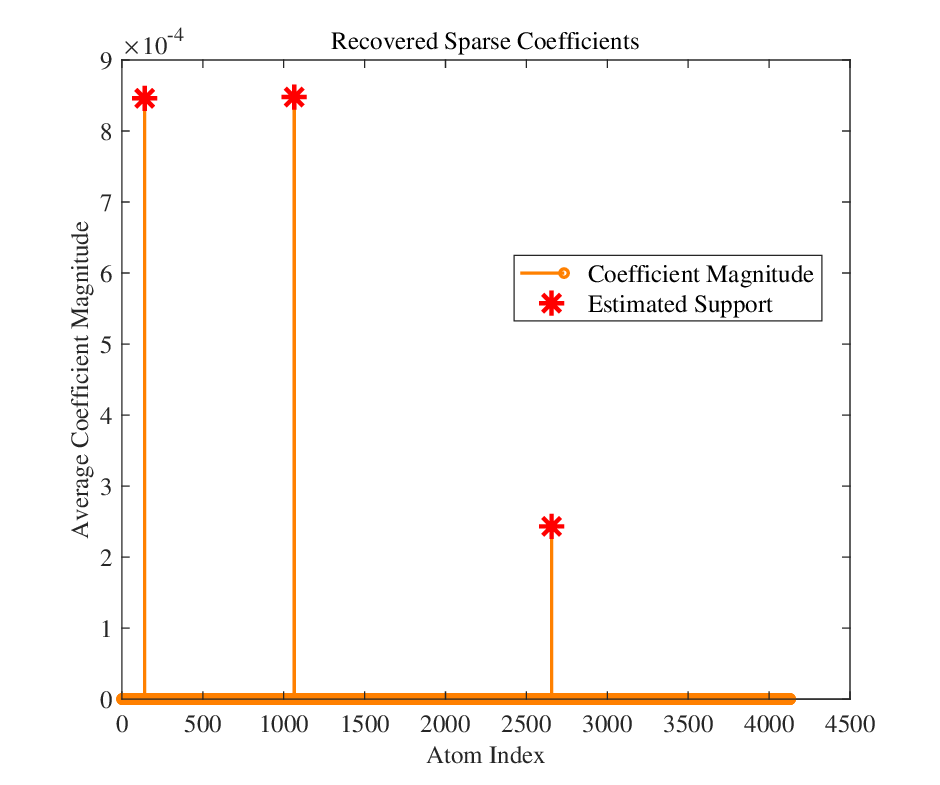}
        \caption{Sparse recovery performance.}
        \label{fig:fig3-a}
    \end{subfigure}
    \hfill 
    \begin{subfigure}[b]{0.48\columnwidth}
        \centering
        \includegraphics[width=\linewidth]{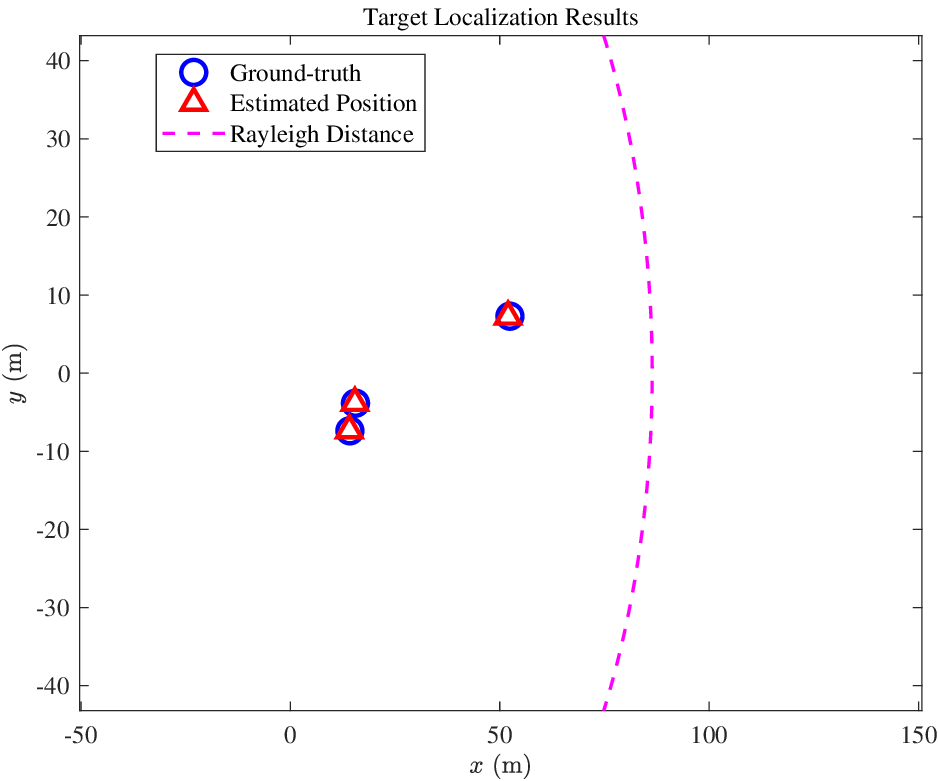}
        \caption{NF localization results.}
        \label{fig:fig3-b}
    \end{subfigure}

    \caption{Visualization of the proposed parameter algorithm.}
    \label{fig:fig-3}\vspace{-2pt}
\end{figure}

We validate the effectiveness of the proposed dictionary design in terms of sparse recovery and demonstrate the overall localization performance. The simulation setup involves $P=3$ targets randomly scattered within the radiative NF region. As shown in Fig.~\ref{fig:fig-3}(\subref{fig:fig3-a}), the recovered sparse coefficients are presented, where the horizontal axis represents the atom indices in the angle-distance dictionary and the vertical axis denotes the average coefficient magnitude. We observe three distinct and prominent peaks, corresponding to three targets.

Furthermore, Fig.~\ref{fig:fig-3}(\subref{fig:fig3-b}) showcases the effectiveness of the proposed algorithm in localizing targets. It plots the true target positions against their estimated counterparts in Cartesian coordinates. The results clearly demonstrate that the estimated positions (marked by red triangles) closely align with the true positions (marked by blue circles), indicating high localization accuracy. Moreover, all targets are accurately identified within the radiative NF region, consistent with the Rayleigh distance.

%
%
%
%

\vspace{-4pt}
\subsection{Performance Analysis of NB-NF MUSIC Method}
\label{sec:V-B}
\subsubsection{Quantification of the Correlation Function}
\begin{figure}
\begin{center}
\includegraphics[width=0.70\linewidth]{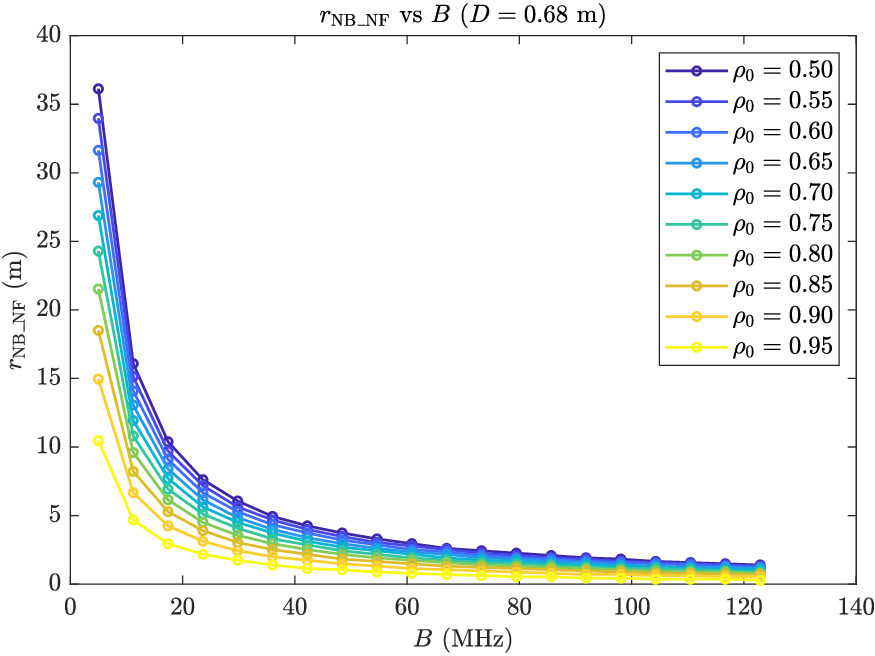}
\end{center}
\caption{The maximum threshold distance $r_{\mathrm{NB-NF}}$ versus system bandwidth $B$ for various correlation thresholds $\rho_0$. }
\label{fig:fig-4}\vspace{-2pt}
\end{figure}
To quantify the limitations of the NB assumption within the WB-NF regime, we evaluate the maximum threshold distance, denoted as $r_{\mathrm{NB-NF}}$. This metric implies the effective upper distance limit beyond which the correlation coefficient $\tilde g_1\left( {{r},{\theta}} \right)$ deteriorates below a predefined reliability threshold $\rho_0$. The simulation considered a fixed target angle of 60 degrees. Fig.~\ref{fig:fig-4} depicts the variation of $r_{\mathrm{NB-NF}}$ against the system bandwidth $B$ (ranging from 5 MHz to 100 MHz), parameterized by varying correlation stringency levels $\rho_0 \in [0.50, 0.95]$. Two primary observations can be drawn from the results:
1) A pronounced inverse dependence between the maximum threshold distance and bandwidth ($r_{\mathrm{NB-NF}} \propto 1/B$) is evident across all correlation levels. For instance, under a high-fidelity requirement of $\rho_0 = 0.90$, the reliable operating range drastically shrinks from approximately 15 m at $B=5$ MHz to less than 1 m at $B=100$ MHz. This indicates that in WB systems, the spatial region where the simple NB model remains valid is negligibly small.
2) This behavior aligns perfectly with Eq.~(\ref{eq:eq-34}), where the array gain is governed by the Dirichlet kernel argument $\Delta k \psi_{n'}(r) \propto B \cdot r$. Consequently, maintaining a fixed correlation $\rho_0$ effectively imposes a constant bandwidth-distance product. As a result, any increase in bandwidth $B$ necessitates a proportional reduction in distance to strictly confine the operating point within the mainlobe.

Physically, $r_{\mathrm{NB-NF}}$ serves as a critical boundary delineating two operating regimes. Beyond this distance ($r > r_{\mathrm{NB-NF}}$), the coherence gain provided by NF effect is approximately nullified by the significant energy defocusing induced by the frequency-dependent phase mismatch. As bandwidth $B$ increases, this WB effect overshadows the focusing gain at progressively shorter distances. Therefore, within the effective distance $r \le r_{\mathrm{NB-NF}}$, the computationally simpler NB-NF model can be adopted to legitimately approximate the complex WB-NF response with high fidelity.

\subsubsection{Validation of the Proposed NB-NF MUSIC Method}
\begin{figure}
\begin{center}
\includegraphics[width=0.99\linewidth]{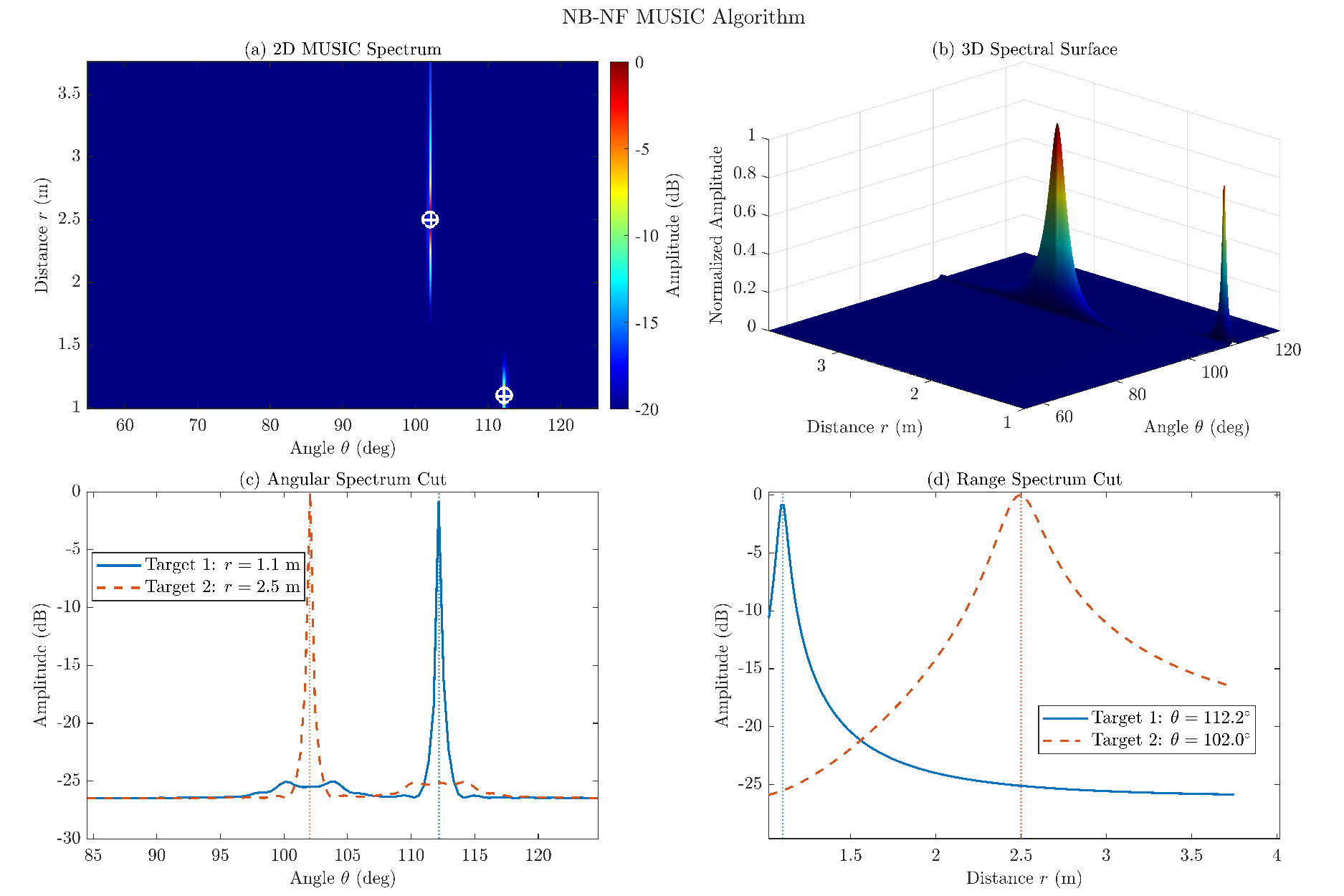}
\end{center}
\caption{Spatial spectrum estimation results of the NB-NF MUSIC algorithm.}
\label{fig:fig-5}\vspace{-2pt}
\end{figure}
Fig.~\ref{fig:fig-5} presents the comprehensive spatial spectrum estimation results of the proposed NB-NF MUSIC algorithm in a dual-target scenario. The simulation setup includes two NF targets randomly located at coordinates $(r_1, \theta_1) = (1.1\,\text{m}, 112.2^\circ)$ and $(r_2, \theta_2) = (2.5\,\text{m}, 102.0^\circ)$. As depicted in the 2D spectrum (Fig.~\ref{fig:fig-5}(a)) and the 3D spectral surface (Fig.~\ref{fig:fig-5}(b)), the algorithm successfully resolves both targets in the spatial domain. However, a significant resolution disparity is observed between the angular and distance dimensions. Specifically, the angular spectrum cut in Fig.~\ref{fig:fig-5}(c) demonstrates extremely sharp spectral peaks with narrow beamwidths, indicating high-resolution angular estimation capabilities. In contrast, the distance spectrum cut in Fig.~\ref{fig:fig-5}(d) reveals substantial spectral broadening and slower decay in the distance domain. This phenomenon, visualized as vertical ridges ((i.e., distance-spreading) in Fig.~\ref{fig:fig-5}(a), suggests that the distance measurement capability of the NB spherical wave model is limited due to the absence of WB time-delay information.


\vspace{-2pt}
\subsection{Performance Analysis of WB-FF MUSIC Method}
\label{sec:V-C}
\subsubsection{Quantification of the Correlation Function}
\begin{figure}
\begin{center}
\includegraphics[width=0.70\linewidth]{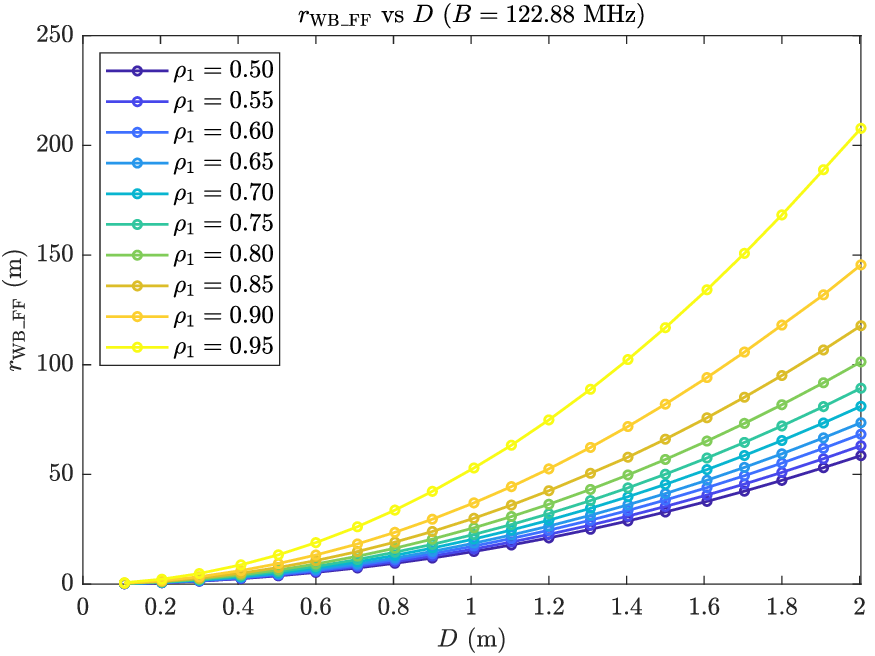}
\end{center}
\caption{The minimum threshold distance $r_{\mathrm{WB-FF}}$ versus system aperture $D$ for various correlation thresholds $\rho_1$. }
\label{fig:fig-6}\vspace{-2pt}
\end{figure}
To quantify the limitations of the FF assumption within the WB-NF regime, we evaluate the minimum threshold distance, denoted as $r_{\mathrm{WB-FF}}$. This metric implies the effective lower distance limit within which the correlation coefficient $\tilde g_2\left( {{r},{\theta}} \right)$ deteriorates below a predefined reliability threshold $\rho_1$. The simulation assumes a fixed bandwidth of $B = 122.88$ MHz based on the derivation in Eq.~(\ref{eq:eq-41}). Fig.~\ref{fig:fig-6} illustrates the relationship between the minimum threshold distance $r_{\mathrm{WB-FF}}$ and the array aperture $D$, parameterized by varying correlation stringency levels $\rho_1 \in [0.50, 0.95]$. Two primary observations can be drawn from the results:
1) The minimum threshold distance exhibits a quadratic growth with respect to the aperture size ($r_{\mathrm{WB-FF}} \propto D^2$). This implies that as the array scale expands, the region where the FF assumption fails expands significantly. 2) The threshold boundary is sensitive to the required correlation $\rho_1$. For a large aperture of $D=1.8$ m, maintaining a high-fidelity correlation of $\rho_1=0.95$ mandates a minimum distance of $168.3$ m. Even relaxing the requirement to $\rho_1=0.85$ only reduces the threshold to approximately $95.1$ m. Conversely, for small apertures (e.g., $D < 0.2$ m), the valid distance starts very close to the array ($r_{\mathrm{WB-FF}} \approx 0$).


\subsubsection{Validation of the Proposed WB-FF MUSIC Method}
\begin{figure}
\begin{center}
\includegraphics[width=0.99\linewidth]{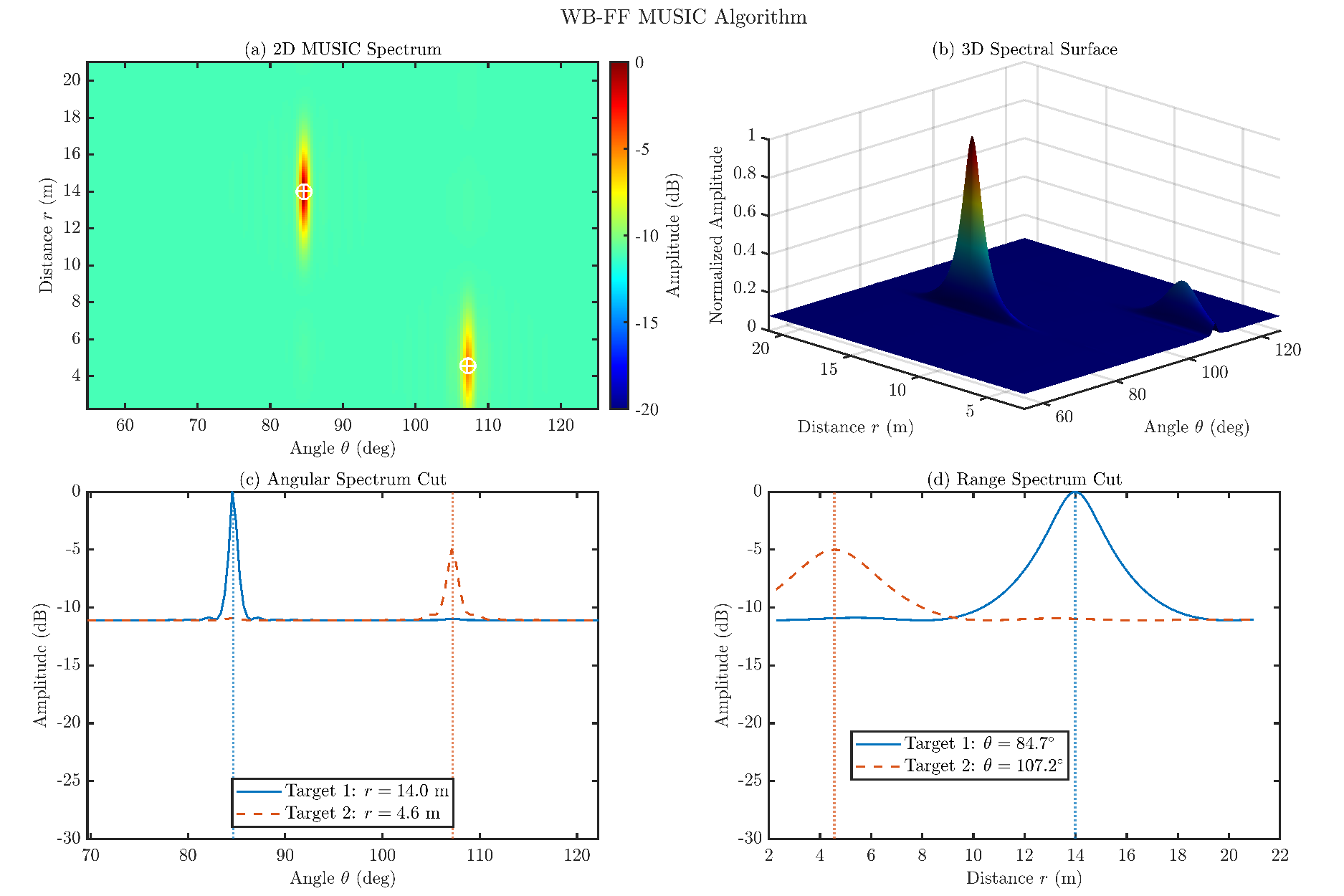}
\end{center}
\caption{Spatial spectrum estimation results of the WB-FF MUSIC algorithm.}
\label{fig:fig-7}\vspace{-2pt}
\end{figure}
Fig.~\ref{fig:fig-7} evaluates the proposed WB-FF MUSIC algorithm using two targets at $(r_1, \theta_1) = (14.0\,\text{m}, 84.7^\circ)$ and $(r_2, \theta_2) = (4.6\,\text{m}, 107.2^\circ)$. As shown in Figs.~\ref{fig:fig-7}(a) and (b), the algorithm successfully resolves both targets by extracting WB time-delay signatures for ranging and utilizing the planar array response for angle estimation. Similar to the NB-NF method, the inherent distance resolution remains lower than the angular resolution, visualized as vertical ridges. More importantly, a severe spatial-dependent performance discrepancy is observed. While the farther target~1 ($14.0\,\text{m}$) exhibits an accurate, sharp spectrum peak near $0\,\text{dB}$, the closer target~2 ($4.6\,\text{m}$) suffers from substantial defocusing. As explicitly shown in Figs.~\ref{fig:fig-7}(c) and~(d), the spectral peak of target~2 degrades to approximately $-5\,\text{dB}$ with noticeable broadening. This severe mismatch occurs because the WB-FF model inherently discards the NF wavefront curvature, which rigorously corroborates the theoretical curvature-induced decorrelation and the critical distance bound $r_{\mathrm{WB-FF}}$ analyzed in Fig.~\ref{fig:fig-6}.

\vspace{-2pt}
\subsection{NMSE vs Distance}
\begin{figure}
\begin{center}
\includegraphics[width=0.72\linewidth]{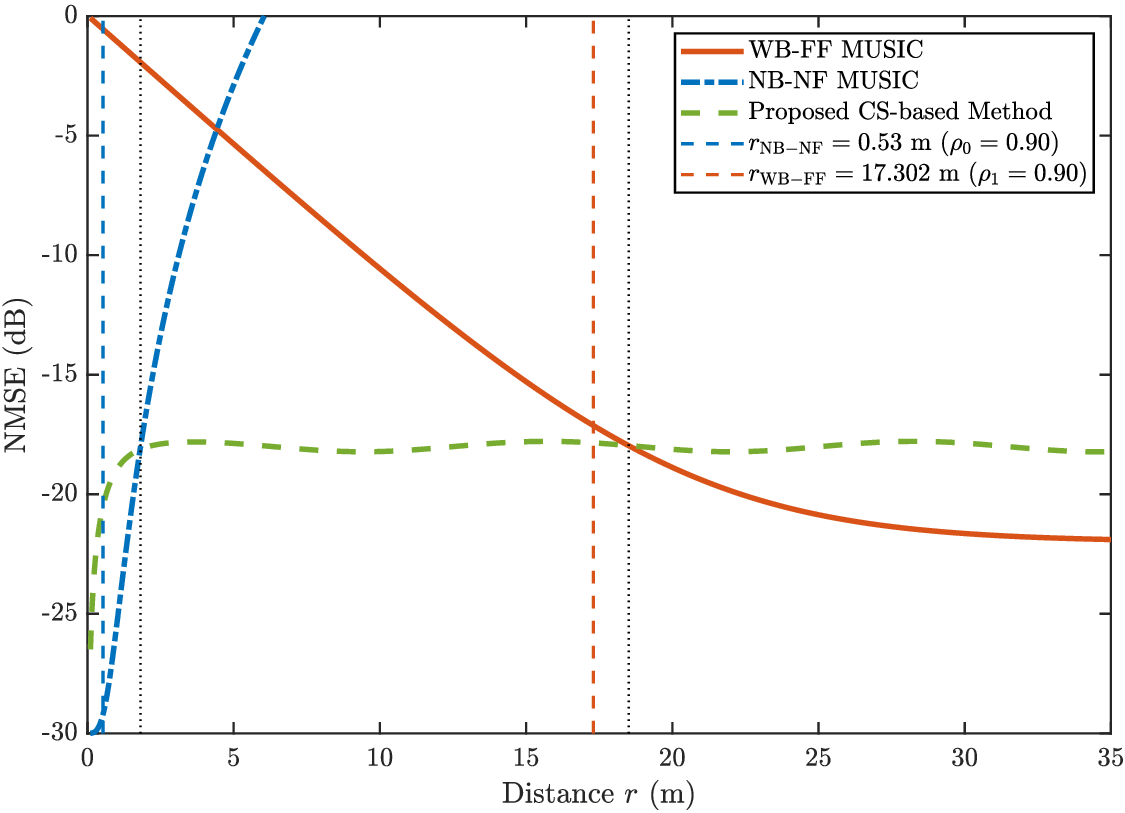}
\end{center}
\caption{NMSE versus distance performance for different algorithms.}
\label{fig:fig-8}\vspace{-5pt}
\end{figure}
In this subsection, we investigate the effective operating regimes of the decoupled approximation models (NB-NF and WB-FF) by evaluating their NMSE against target distance. This analysis aims to identify the specific distance intervals where these low-complexity substitutions remain viable.

As illustrated in Fig.~\ref{fig:fig-8}, although the geometric Rayleigh distance is $R_r \approx 86.4$~m, both decoupled methods exhibit highly restricted ``sweet spots''. For the NB-NF method (blue curve), distance sensing precision relies on spherical wavefront curvature. As established analytically, its spatial phase coherence ($\rho \ge 0.9$) holds only within $r \le r_{\mathrm{NB-NF}} = 0.53$~m. Instead of failing abruptly at this conservative theoretical boundary, the spatial correlation degrades progressively, allowing the NB-NF MUSIC method to exploit residual wavefront curvature for viable estimation slightly beyond this limit. However, the NMSE escalates at a rapid rate (proportional to $(r/D)^4$) as curvature sensitivity diminishes. At the first empirical intersection ($r \approx 1.8$~m), the noise floor dominates the weakened curvature features, causing the NB-NF performance to plummet below the proposed CS-based method.

Conversely, the WB-FF MUSIC approach (red curve) leverages the WB envelope delay ($B = 122.88$~MHz) for long-range targets. While severely degraded by uncompensated quadratic phase errors at close-in distances, its performance progressively stabilizes once the distance exceeds the theoretical boundary of $r_{\mathrm{WB-FF}} = 17.302$~m ($\rho = 0.9$), where the planar wavefront assumption becomes tolerable. Nevertheless, since the residual spherical curvature still incurs a marginal phase penalty at this exact threshold, the empirical WB-FF curve requires a slightly extended distance to fully surpass the CS baseline. Consequently, the actual intersection occurs at $r \approx 18.5$~m, beyond which the approach recovers its bandwidth-dependent distance sensing capability.

Crucially, the extensive region between $1.8$~m and $18.5$~m manifests as a severe ``gray zone'' where neither decoupled model yields acceptable accuracy. In this transitionary phase, the target is neither sufficiently close for reliable curvature extraction nor sufficiently far to ignore the NF phase mismatch. This observation justifies the necessity of our proposed unified sensing framework, which is capable of bridging this gap and ensuring reliability across the entire sensing range.

%
\vspace{-4pt}
\section{Conclusion}
This paper investigated target sensing within WB-NF ISAC systems for 6G networks. To address the limitations of conventional WB-FF and NB-NF assumptions, we proposed a unified CS-based parameter estimation algorithm. Specifically, we formulated an exact WB-NF sparse matrix and designed a novel low-coherence angle-distance dictionary, which significantly reduced computational complexity without sacrificing resolution. Furthermore, to perform a decoupled NF and WB effect analysis, we developed two MUSIC-based benchmark schemes. By evaluating the estimation performance against these baselines, we systematically analyzed the isolated physical effects and identified the distinct effective operating regimes.

\bibliographystyle{IEEEtran}
\bibliography{reference}



\end{document}